\documentclass[a4paper,11pt]{article}
\pdfoutput=1 

\usepackage{jcappub} 
\usepackage{xcolor}
\usepackage[T1]{fontenc} 
\usepackage{bm}
\usepackage{epsfig}
\usepackage[lofdepth,lotdepth,caption=false]{subfig}
\usepackage{graphicx}

\usepackage{float}
\usepackage{amsmath,amssymb}
\usepackage{textcomp}
\usepackage{gensymb}
\usepackage[utf8]{inputenc}
\usepackage{setspace}
\usepackage{hyperref}
\usepackage[normalem]{ulem}
\usepackage{pdflscape}
\usepackage{multirow}

\def\beq{\begin{equation}}
\def\enq{\end{equation}}
\def\bea{\begin{eqnarray}}
\def\ena{\end{eqnarray}}

\begin{document}
\begin{flushright}
LAPTH-031/20
\end{flushright}

\title{Where do IceCube neutrinos come from? Hints from the diffuse gamma-ray flux}
\author[a]{Antonio Capanema,}
\author[a]{Arman Esmaili,}
\author[b]{Pasquale Dario Serpico}
\emailAdd{antoniogalvao@aluno.puc-rio.br}
\emailAdd{arman@puc-rio.br}
\emailAdd{serpico@lapth.cnrs.fr}
\affiliation[a]{Departamento de F\'isica, Pontif\'icia Universidade Cat\'olica do Rio de Janeiro, Rio de Janeiro 22452-970, Brazil}
\affiliation[b]{LAPTh, Univ. Grenoble Alpes, USMB, CNRS, F-74000 Annecy, France}

\abstract{Despite the spectacular discovery of an astrophysical neutrino flux by IceCube in 2013, its origin remains a mystery. Whatever its sources, we expect the neutrino flux to be accompanied by a comparable gamma-ray flux. These photons should be degraded in energy by electromagnetic cascades and contribute to the diffuse GeV-TeV flux precisely measured by the Fermi-LAT. Population studies have also permitted to identify the main classes of contributors to this flux, which at the same time have not been associated with major neutrino sources in cross-correlation studies. These considerations allow one to set constraints on the origin and spectrum of the IceCube flux, in particular its low-energy part. We find that, even accounting for known systematic errors, the Fermi-LAT data exclude to at least 95\% C.L. any extragalactic transparent source class, irrespective of its redshift evolution, if the neutrino spectrum extends to the TeV scale or below. If the neutrino spectrum has an abrupt cutoff at $\sim10$~TeV, barely compatible with current observations, the tension can be reduced, but this way out requires a significant modification to the current understanding of the origin of the diffuse extragalactic gamma-ray flux at GeV energies. In contrast, these considerations do not apply if a sizable fraction of IceCube data originates within the Galactic halo (a scenario however typically in tension with other constraints) or from a yet unidentified class of ``opaque'' extragalactic emitters, which do not let the high-energy gamma rays get out.}
\maketitle
\date{\today}


\section{\label{sec:intro}Introduction}

The era of \textit{high-energy neutrino astronomy} has started with the intriguing IceCube detection of a diffuse neutrino flux covering the energy range of $\sim10$~TeV to several PeV~\cite{Aartsen:2013bka,Aartsen:2013jdh,Aartsen:2015rwa,Aartsen:2020aqd}. For the first time, we are observing the Universe through extremely energetic and penetrative messengers which can bring us information from cosmological distances and/or from the interior of opaque sources. Extragalactic astronomy in this energy range seems to be possible only via neutrinos: The photon attenuation length decreases with an increase in energy, being limited to Galactic distances at $\sim$~PeV, while stable charged particles (protons/nuclei) deflect en route to the Earth and lose any directional information. Current theories also suggest that Galactic cosmic rays dominate the flux at these energies, and the cosmic ray extragalactic sky may be inaccessible due to the ``magnetic horizon'' effect, see e.g.~\cite{Mollerach:2013dza}.

If we think of how wildly different the sky looks in visible light and in the X-ray band, for instance, we can easily conceive that new classes of astrophysical objects can pop up in the unexplored energy range of IceCube, potentially changing our understanding of high energy astrophysical processes and acceleration mechanisms. Also, astrophysics and cosmology presents us with many puzzles, such as the nature of dark matter, unexplained within the standard model of particle physics. As a result of the null searches at high-energy colliders and in precision experiments, the argument that the answers to these mysteries is to be found at the electroweak scale appears less and less convincing. Perhaps, new astrophysical windows can offer new insights and opportunities. 

The source(s) of the observed diffuse neutrino flux by IceCube is(are) yet unknown. The traditional angular correlation analysis with various catalogs of known astrophysical objects did not lead to any significant association: For instance, blazars seem to contribute to the observed diffuse neutrino flux by $\lesssim\mathcal{O}(10\%)$~\cite{Huber:2019lrm,Aartsen:2016lir}, the lack of clustering and/or multiplets in $\mu$-track events excludes strong point sources~\cite{Aartsen:2018ywr}, and time correlation analysis rejects GRB~\cite{Aartsen:2016qcr} and chocked-jet supernovae~\cite{Senno:2017vtd,Esmaili:2018wnv} contributions. In this context, a multi-messenger approach is of crucial importance to finding clues about the source(s). In particular, any source of high energy neutrinos is unavoidably associated to a $\gamma$-ray counterpart at similar energies. However, the propagation of such $\gamma$-rays from the source to the Earth is less trivial: far from crossing a transparent medium, they initiate electromagnetic cascades by successive pair-production on and inverse-Compton scattering off the cosmic microwave background (CMB) and extragalactic background light (EBL)~\cite{Stecker:2006ips,Franceschini:2008eob,Kneiske:2010alf,Dominguez:2010bv,Stecker:2016aed}, resulting in a diffuse flux of $\gamma$-rays with energies $\lesssim1$~TeV. Thanks to Fermi-LAT, its measurement of the extragalactic $\gamma$-ray background (EGB) and determination of the isotropic diffuse $\gamma$-ray background (IGRB)~\cite{Ackermann:2014usa}, we can constrain (or detect) contributions from the electromagnetic cascade and thus probe scenarios for the sources of high energy neutrinos. 

By requiring that the flux from electromagnetic cascades should not overshoot the IGRB, severe constraints have been derived in~\cite{Murase:2013rfa,Murase:2015xka,Kistler:2015ywn,Bechtol:2015uqb,Chang:2014hua,Tamborra:2014xia,Chang:2016ljk,Xiao:2016rvd}; in~\cite{Ando:2015bva} the cross-correlation of IGRB data with galaxy distributions has been used. However, in order to make the tension quantitative and to be statistically consistent, one has to derive the constraints from the EGB by taking all the uncertainties and contributions into account; especially paying attention to the fact that a large fraction of the EGB is nowadays attributed , with a growing degree of confidence, to the unresolved component of known classes of objects. 
Employing this method, in~\cite{Capanema:2020rjj} it has been recently  shown that the observed diffuse neutrino flux in the 6-years cascade data set of IceCube, which extends down to $\sim 10$~TeV, is in $\gtrsim3\sigma$ tension with the EGB data for transparent source classes with $z$-distribution resembling the star formation rate (SFR). The tension worsens to $\gtrsim5\sigma$ if the diffuse neutrino flux extends down to $\sim1$~TeV. Given the impact of these conclusions, we deem important to further investigate how {\it robust and generic} these tensions are and what is the information one can draw from it. This is the subject of this article. 

The paper is structured as follows: in section~\ref{sec:prem}, we review the neutrino (section~\ref{sec:ICdata}) and $\gamma$-ray (section~\ref{sec:EGBdata}) data sets, and recall the theoretical links between the two (section~\ref{sec:nugamma}). We also present our data analysis approach, which includes some refinements compared to the one adopted in~\cite{Capanema:2020rjj}. In section~\ref{sec:rob}, we inspect the robustness of the bounds reported in~\cite{Capanema:2020rjj}. In detail, we update the expected contributions to the EGB of sources following either the SFR evolution (section~\ref{sec:sfr}) or the blazar one (section~\ref{sec:bl}), improving upon the analysis method and considering the uncertainties in EGB data from Galactic foreground modeling. We anticipate that this detailed analysis shows a persisting tension: for a diffuse neutrino flux extending down to $\sim1$~TeV, this is quantitatively significant even if only the high energy part ($>10$~GeV) of EGB data is taken into account. Only if the neutrino flux is strongly suppressed below $\sim 10$~TeV, we find that it is possible to reduce the tension to milder values; however, this would imply a failure to interpret the low energy part of EGB data as sourced by radio galaxies and blazars within current models. In the remaining sections of the paper, we extend the analysis to the most obvious alternatives: i) Sources at very high redshifts, as for instance related to first astrophysical objects responsible for the reionization era (section~\ref{sec:highz}); ii) Extremely close sources (section~\ref{sec:lowz}). For different reasons, neither offers a fully satisfactory way to escape the bounds, as we detail in the respective sections. In section~\ref{concl} we provide a summary discussion of our results, some of their implications and conclude.

\section{\label{sec:prem}Neutrino and $\gamma$-ray ray data and their connection}
In this section we briefly summarize the neutrino and $\gamma$-ray data sets used in our analysis, respectively, in sections~\ref{sec:ICdata} and \ref{sec:EGBdata}. Section~\ref{sec:EGBdata} also describes the established contributions to the EGB from various astrophysical classes of sources. Section~\ref{sec:nugamma} outlines the phenomenological implications of the theoretical connection between these data sets, thus setting the rationale for the multi-messenger approach pursued in this paper. 

\subsection{\label{sec:ICdata}IceCube's neutrino data sets}

The first observation of cosmic neutrinos by the IceCube detector, consisting of two $\sim$~PeV events in less than two years of data, has been reported more than seven years ago~\cite{Aartsen:2013jdh}. Since then, the sample has considerably grown. The data are usually classified based on the event-topology ($\mu$-tracks or cascades) and/or the position of interaction vertex (inside or outside the fiducial volume of detector). There are thus three data sets of IceCube's neutrinos in recent publications: \textit{i}) The high energy starting events (HESE) data set comprising both $\mu$-track and cascade events with the interaction vertex inside the detector; the latest publication includes 7.5 years of data~\cite{Schneider:2019ayi}. \textit{ii}) The through-going $\mu$-track (TG) data set consisting of all the $\mu$-track events, independently of the interaction vertex position, whose latest update includes 9.5 years of data~\cite{Stettner:2019tok}. \textit{iii}) The data set of cascade (or shower-like) events, recently including 6 years of data~\cite{Aartsen:2020aqd}.  

Assuming equal abundances of neutrinos and antineutrinos of all flavors, each neutrino data set can be conveniently fit in terms of the normalization $\Phi_{\rm astro}$ and spectral index $s_{\rm ob}$ of a (one-flavor neutrino) power-law, anchored to the energy 100 TeV, as:
\begin{equation}\label{eq:icflux}
\Phi_\nu = \frac{{\rm d}\phi_\nu}{{\rm d}E_\nu} = 10^{-18}\, \Phi_{\rm astro} \left( \frac{E_\nu}{100~{\rm TeV}}\right)^{-s_{\rm ob}}\quad,\qquad {\rm for}~~ E_\nu \geq E_{\rm thr}~, 
\end{equation}
where $\Phi_{\rm astro}$ is in units of ${\rm GeV}^{-1}{\rm cm}^{-2}{\rm s}^{-1}{\rm sr}^{-1}$. The threshold energy, $E_{\rm thr}$, which depends on the background rejection efficiency, is the minimum energy considered in the fit\footnote{It is worth clarifying that the threshold energy reported  by IceCube does not indicate the lowest energy observed event in each data set, which is always lower than $E_{\rm thr}$. In fact, the reported values for threshold energy originate from an optimization of the fit to the data. The analyses in~\cite{Capanema:2020rjj} and this paper hopefully highlight the importance of the low-energy part of the data, motivating dedicated searches for the lowest energy events.}. Table~\ref{tab:nudata} summarizes the measured values of parameters in Eq.~(\ref{eq:icflux}) for the three IceCube data sets.  

\begin{table}[t!]
    \centering
    \begin{tabular}{|c|c|c|c|}
    \hline
     & $\Phi_{\rm astro}$ & $s_{\rm ob}$ & $E_{\rm thr}$ \\ \hline
    HESE & $2.15^{+0.49}_{-0.15}$ & $2.89^{+0.2}_{-0.19}$ & 60 TeV \\ \hline
    TG & $1.44^{+0.25}_{-0.24}$ & $2.28^{+0.08}_{-0.09}$ & 119 TeV \\ \hline
    Cascade & $1.66^{+0.25}_{-0.27}$ & $2.53^{+0.07}_{-0.07}$ & 16 TeV \\ \hline
    \end{tabular}
    \caption{The measured values of parameters in Eq.~(\ref{eq:icflux}) for the three IceCube data sets. The errors are at $1\sigma$. The normalization $\Phi_{\rm astro}$, with units ${\rm GeV}^{-1}{\rm cm}^{-2}{\rm s}^{-1}{\rm sr}^{-1}$, is for one flavor and assuming $1:1:1$ flavor ratio at Earth.}
    \label{tab:nudata}
\end{table}

\subsection{\label{sec:EGBdata}EGB data and contributions}

To be faithful to its name, the extragalactic $\gamma$-ray background (EGB) should be the sum of all the $\gamma$-rays emitted from  sources outside the Milky Way. In practice, however, at least within the Fermi-LAT collaboration, one defines the EGB ``operationally'', as the total {\it bona fide} $\gamma$-ray flux in the detector, minus the estimated foreground of diffuse Galactic emission (above Galactic latitude $|b|>20^\circ$) and Solar emission. This includes the (quasi-isotropic) diffuse $\gamma$-ray background (IGRB) and any resolved or unresolved sources, making the EGB largely independent from the performance of the instrument~\footnote{But for the rejection of the misidentified cosmic ray background.}, but dependent on the modeling of the Galactic diffuse/extended backgrounds. The latest measurement of EGB by the Fermi-LAT reports the EGB flux in the energy range 100 MeV to 820 GeV~\cite{Ackermann:2014usa}. The cosmic ray interactions with interstellar gas and the inverse-Compton scattering off the interstellar radiation field, which are responsible the Galactic emission foreground, have been modeled in~\cite{Ackermann:2014usa} by taking into account various scenarios for both electron and nuclei injection and diffusion-loss propagation in our Galaxy. The EGB flux has been reported for three foreground models, labeled \textbf{A}, \textbf{B} and \textbf{C} in~\cite{Ackermann:2014usa}, which we refer the reader to for further details. In the following, we take the variability among those models to represent the systematic error in the ``cleaning'' of the Galactic emission foreground when extracting the EGB. 

Based on current understanding, the bulk of the EGB is due to the sum of resolved and unresolved extragalactic point-like object emission. 
For the (averaged) resolved point-like sources contributing to the EGB, we use updated evaluations with respect to~\cite{Capanema:2020rjj}. Three main contributions to the EGB from resolved sources have been contemplated in the literature as (in the order of decreasing percentage contribution to the whole energy range of EGB): i) The blazar~\footnote{This class contains both BL Lac objects and flat spectrum radio quasars, a sub-division sometimes taken into account and modeled independently in gamma-ray studies, see e.g.~\cite{DiMauro:2013zfa}.} $\gamma$-ray contribution; for its luminosity function we take the ``luminosity-dependent density evolution'' model from~\cite{Qu:2019zln}, but we will also compare our benchmark with the alternative modeling of~\cite{DiMauro:2015tfa}. ii) Radio galaxies, or misaligned active galactic nuclei (mAGN), which we model according to~\cite{DiMauro:2013xta}. iii) Star-Forming Galaxies (SFG), contributing to the EGB at $\sim5\%$ level, for which  we use the results of~\cite{Ajello:2020zna}. Note that the leading unresolved contribution to the EGB from unresolved Galactic objects is estimated to be from Galactic millisecond pulsars, amounting to $\lesssim0.1\%$~\cite{Calore:2014oga}. This is well below the difference among the models \textbf{A}, \textbf{B} and \textbf{C} and, while we take it into account for completeness, it is irrelevant for the following analysis.

\begin{figure}[t!]
\centering
\includegraphics[width=0.8\textwidth]{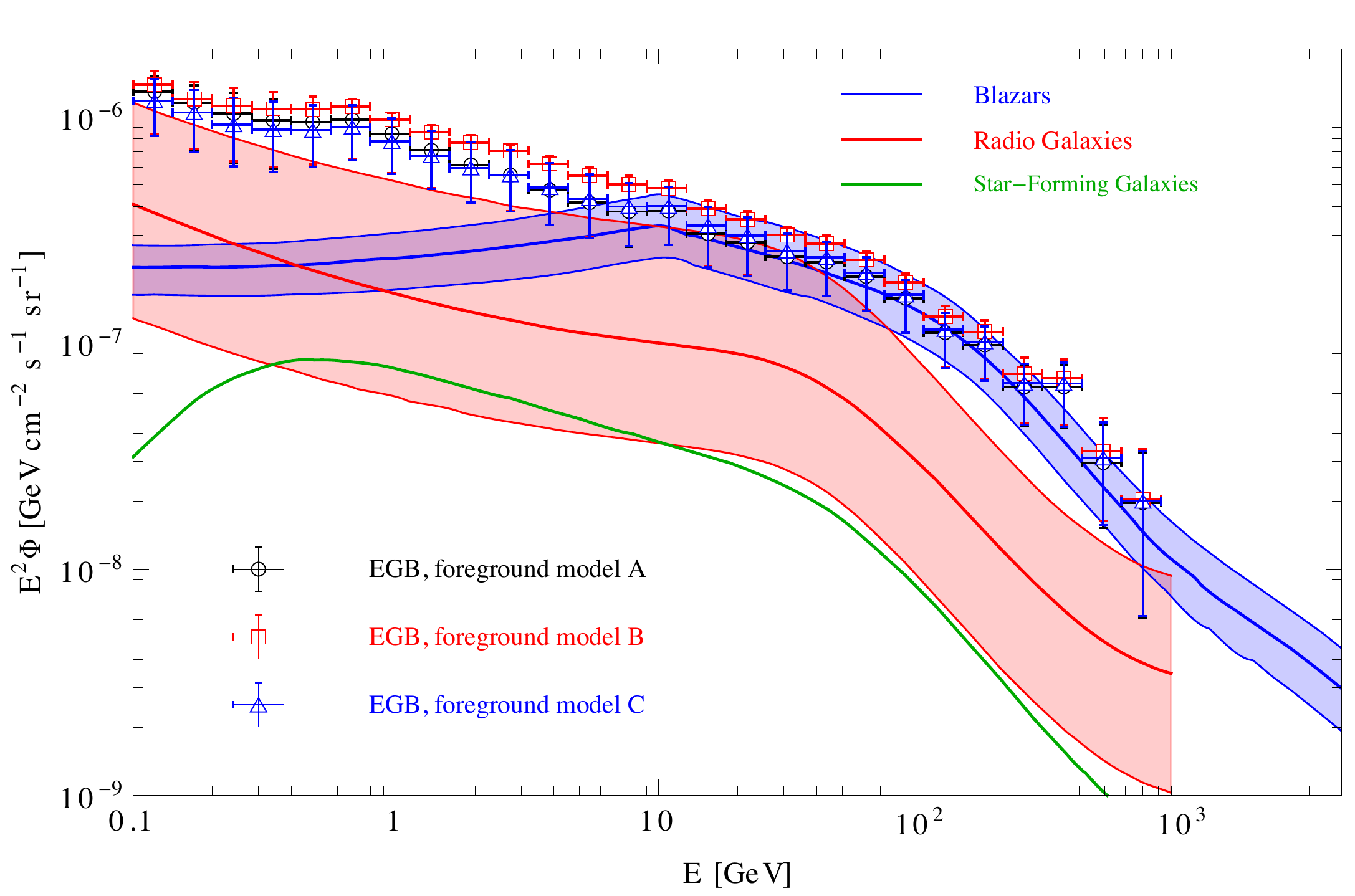}
\caption{\label{fig:egb}The EGB data, for three different foreground modelings, taken from~\cite{Ackermann:2014usa}. The $1\sigma$ error bars include statistical, systematic and foreground uncertainty errors (added in quadrature). The blue, red and green curves show the contributions to EGB from blazars~\cite{Qu:2019zln}, mAGNs~\cite{DiMauro:2013xta} and SFG~\cite{Ajello:2020zna}, respectively. The shaded regions show the $1\sigma$ uncertainty. The contributions from SFG and pulsars (which is very small and below the flux range shown in this figure) are fixed in our analysis, so the uncertainty is not shown.}
\end{figure}

The three data sets of EGB (for different foreground modelings) and the above-cited contributions to EGB are shown in Figure~\ref{fig:egb}. The error bars of EGB data include statistical, systematic and foreground uncertainties added in quadrature. The shaded regions for blazars and mAGN contributions show the estimated 68\% CL uncertainty (which we will also loosely refer to as 1$\sigma$). Since the contributions from SFG and, a fortiori, millisecond pulsars are very small, we keep them fixed in our analysis.      

Blazars alone explain fairly well the EGB spectrum above $\sim 10$~GeV: this conclusion is consistent with independent analyses involving the angular power spectrum and the pixel statistics of the $\gamma$-ray data, see~\cite{Manconi:2019ynl}. In conjunction with the other components, notably mAGNs, blazars provide a decent fit of the EGB data over the whole Fermi-LAT energy range. In order to quantify statistically the fit, we perform a $\chi^2$ analysis whose simplified form of $\chi^2$ function writes:
\begin{equation}\label{eq:chi2-egb}
\chi^2 = \min_{\alpha_j} \left\{\left[ \sum_i \frac{\left(F_{i}^{\rm EGB}-\sum_j\alpha_j F_{i}^{j}\right)^2}{\sigma_i^2} \right] +\sum_j \left[ \frac{(\alpha_j -1)^2}{\varsigma_j^2} \right]\right\}\,.    
\end{equation}
In this formula, $F_{i}^{\rm EGB}$ is the intensity of EGB in the $i-$th bin of energy, with uncertainty $\sigma_i$; the terms $F_{i}^{j}$ are the contributions of the $j-$th class of sources to the energy bin $i$. The nuisance parameters $\alpha_j$ take into account the uncertainties in the contributions of the different classes of sources, and in the second square brackets of Eq.~(\ref{eq:chi2-egb})---the so-called pull terms---their assumed Gaussian variance has been denoted via $\varsigma_j^2$. 

Compared to the relatively simple Eq.~(\ref{eq:chi2-egb}), we actually use a slightly modified $\chi^2$ form accounting for three facts: i) As can be seen from Table~\ref{tab:egbdata} in appendix \ref{sec:app}, which reports values of $F_{i}^{\rm EGB}$ and $\sigma_i$ in $\left[{\rm cm}^{-2}\,{\rm s}^{-1}\,{\rm sr}^{-1}\right]$, the total errors of EGB data are asymmetric, mainly due to the asymmetric errors from foreground modeling. We have thus modified the $\chi^2$ function in Eq.~(\ref{eq:chi2-egb}) to take this fact into account. ii) The meta-parameters $\alpha_j$ have been allowed to depart from unity (and thus contribute to the pull term) only for blazar and mAGN, since these are by far the dominant contributors. We use $\varsigma=0.34,2.02$ for blazar and mAGN, respectively. iii) We consider two different types of pull-terms in Eq.~(\ref{eq:chi2-egb}): besides the Gaussian term reported, we also test a log-normal form. While the Gaussian pull-term is more common in literature, the log-normal type is a more realistic description for order-of-magnitude uncertainties such as the one affecting mAGN. The results we obtain in this paper are almost the same for both types of the pull-terms. 

\begin{figure}[t!]
\centering
\subfloat{
\includegraphics[width=0.33\textwidth]{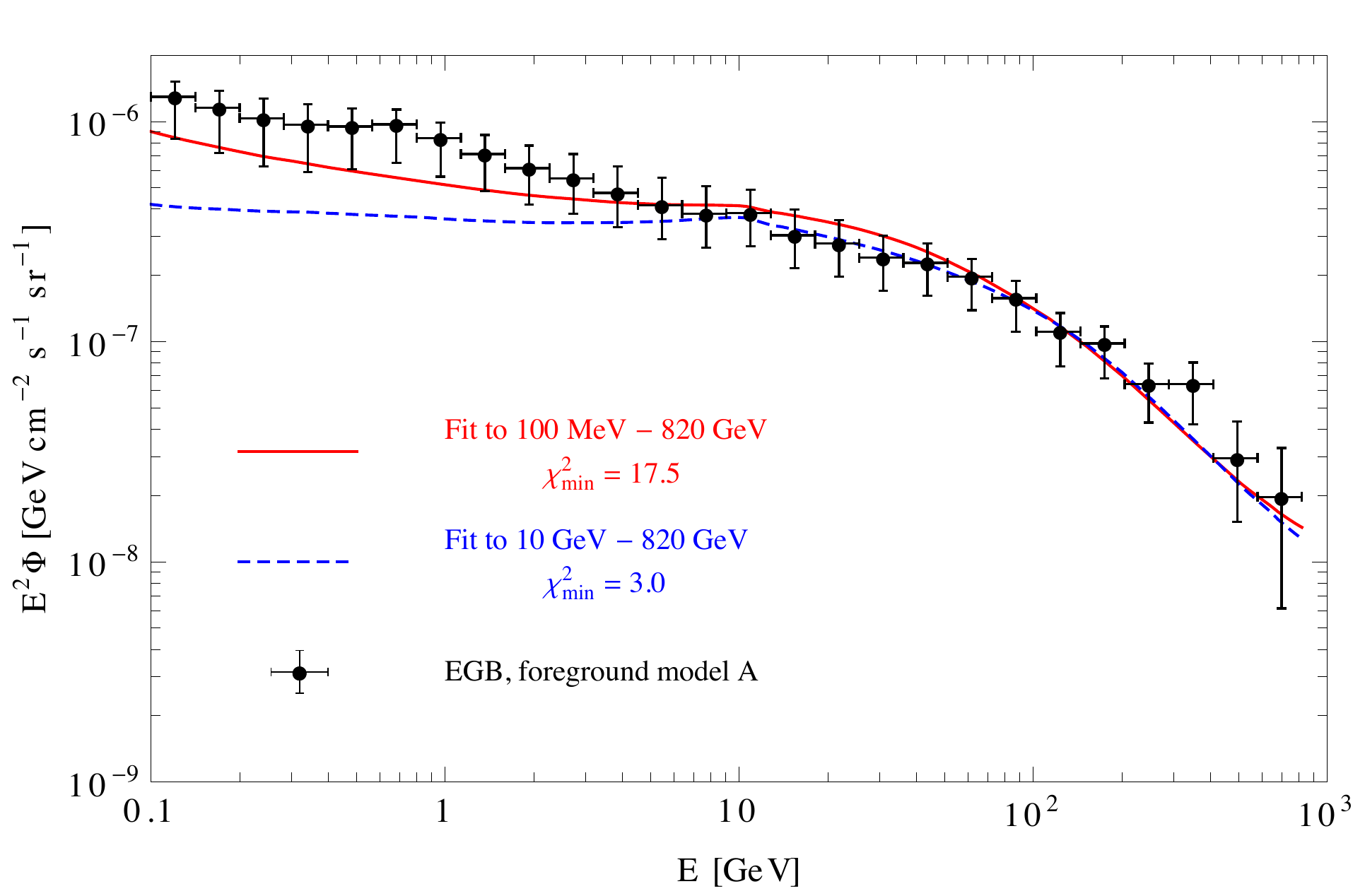}
\label{fig:modelA}
}
\subfloat{
\includegraphics[width=0.33\textwidth]{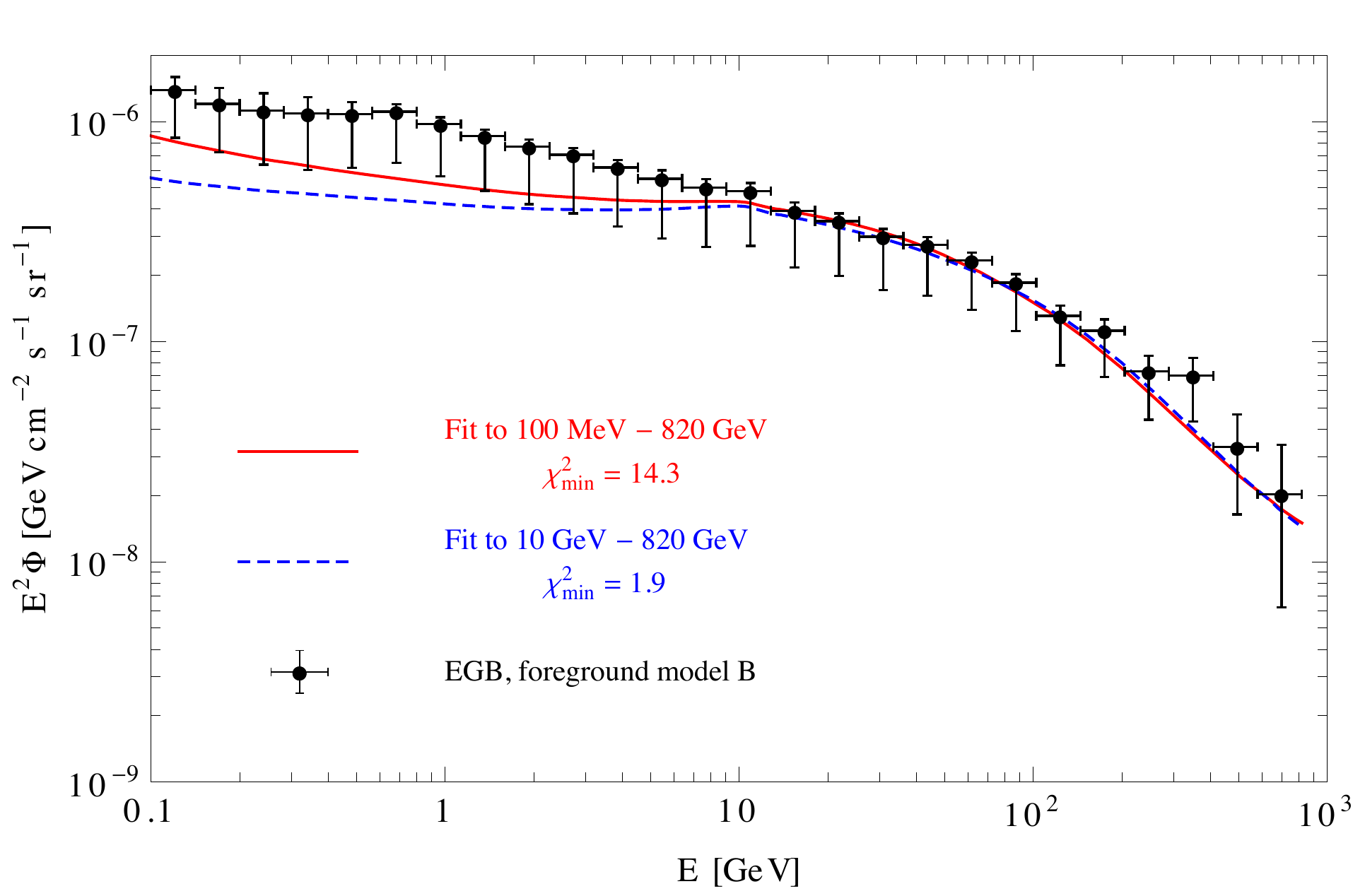}
\label{fig:modelB}
}
\subfloat{
\includegraphics[width=0.33\textwidth]{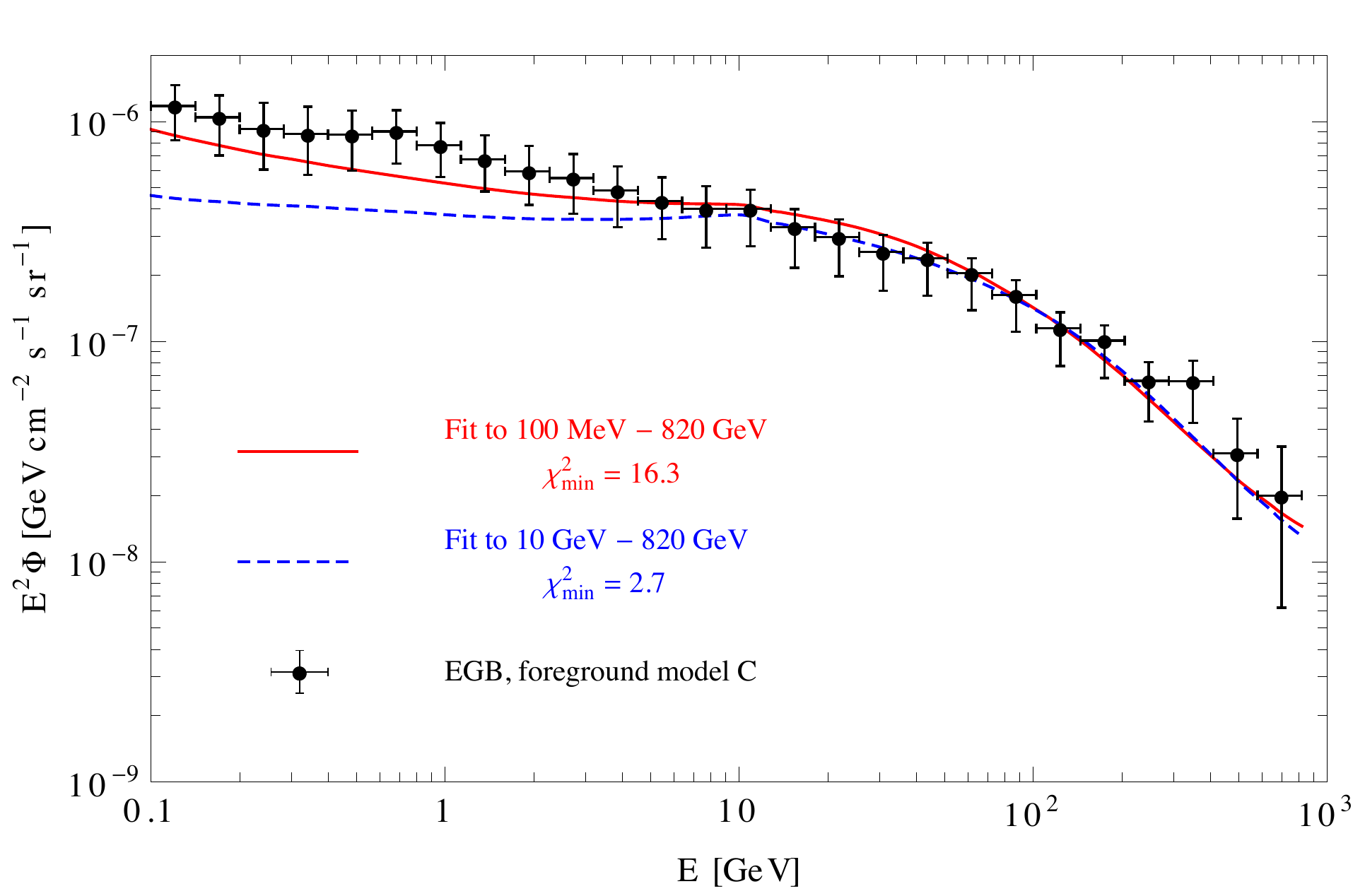}
\label{fig:modelC}
}
\caption{\label{fig:egb-fit}Fits to the EGB data for three modelings of foreground. In each panel, the red-solid and blue-dashed curves show the fit to data in the energy range (100 MeV-820 GeV) and (10 GeV-820 GeV), respectively.}
\end{figure}

Finally, we consider two EGB energy ranges in our analyses: the whole energy range of 100 MeV to 820 GeV, and the high energy part (10-820) GeV exclusively. In the former analysis, the summation in Eq.~(\ref{eq:chi2-egb}) runs over $i=1$ to 26, while in the latter analysis from $i=14$ to 26. The panels of Figure~\ref{fig:egb-fit} show the fits to three modelings of EGB data. The minimum of $\chi^2$ in Eq.~(\ref{eq:chi2-egb}) for each analysis is shown in the legends. The blazar, mAGN and SFG contributions provide a good fit to the whole EGB data for all the three foreground modelings: the minimum of $\chi^2$ varies between $\sim14-17$ among the three of them. Considering the energy range $>10$~GeV, the contributions in fact overfit the EGB data, with $\chi^2_{\rm min}\sim2-3$. We stress however that these figures should not be over-interpreted, since the Fermi-LAT experiment does not report error correlations, which could be important as already revealed in other domains of precision astroparticle physics (see for instance~\cite{Genolini:2019ewc,Boudaud:2019efq}.).

\subsection{\label{sec:nugamma}Neutrino-Gamma connection}

Any population of source(s) with a specific cosmic evolution which can account for the observed diffuse flux by IceCube must be accompanied by a $\gamma$-ray flux. The $\gamma$-rays initiate electromagnetic cascades during their propagation to the Earth and contribute to the diffuse extragalactic $\gamma$-ray background, as discussed in section~\ref{sec:EGBdata}. In the following we briefly discuss this interrelation between neutrinos and $\gamma$-rays, which will be used  in the following sections to constrain possible source populations of neutrinos.

Let us consider a typical source with the (all flavor) neutrino yield $\frac{{\rm d}N_\nu}{{\rm d}\varepsilon_\nu} (\varepsilon_\nu)$ in $\left[{\rm GeV}^{-1}\,{\rm s}^{-1}\right]$, where $\varepsilon_\nu$ is the energy of neutrinos at production. The differential diffuse neutrino flux at the Earth, $\frac{{\rm d}\phi_{\nu}}{{\rm d}E_\nu}(E_\nu)$ in $\left[{\rm GeV}^{-1}\,{\rm cm}^{-2}\,{\rm s}^{-1}\,{\rm sr}^{-1}\right]$, from a population of these sources with cosmic evolution $\mathcal{F}(z)$ in $\left[ {\rm cm}^{-3}\right]$ is given by
\begin{equation}\label{eq:nudiff}
\frac{{\rm d}\phi_{\nu}}{{\rm d}E_\nu} (E_\nu) = \frac{1}{4\pi} \int {\rm d}z\, \frac{{\rm d}\mathcal{V}_c}{{\rm d}z}\, \mathcal{F}(z) \,\frac{1}{4\pi d_c^2}\, \frac{{\rm d}N_\nu}{{\rm d}\varepsilon_\nu} \left[(1+z)E_\nu\right]~,      
\end{equation}
where $E_\nu$ is the neutrino energy at Earth, $\mathcal{V}_c$ is the comoving volume and $d_c$ is the comoving distance. 

The production yield $\frac{{\rm d}N_\nu}{{\rm d}\varepsilon_\nu}$ should be chosen such that the diffuse flux $\frac{{\rm d}\phi_{\nu}}{{\rm d}E_\nu}$ provides a decent fit to the IceCube data. Taking a {\it minimal} approach, we assume 
\begin{equation}\label{eq:nuspec}
\frac{{\rm d}N_\nu}{{\rm d}\varepsilon_\nu} = \begin{cases} 
A & \varepsilon_\nu < \varepsilon_{\rm br} \\
A(\varepsilon_\nu/\varepsilon_{\rm br})^{-s_h} & \varepsilon_{\rm br}\leq \varepsilon_\nu \leq 10\,{\rm PeV} \\
0 & \varepsilon_\nu > 10\,{\rm PeV} 
\end{cases}~,
\end{equation}
where $s_h$ is the energy index and the break energy, $\varepsilon_{\rm br}$, introduces a low-energy cutoff on the spectrum, while the high-energy cutoff of $10$~PeV is motivated by the non-observation of neutrinos with higher energies in IceCube. The constant $A$ parameterizes the normalization of neutrino yield. The allowed ranges for the free parameters in Eq.~(\ref{eq:nuspec}), that are $s_h$, $\varepsilon_{\rm br}$ and $A$, can be found by comparing the diffuse flux in Eq.~(\ref{eq:nudiff}) with IceCube's data sets in Table~\ref{tab:nudata}. The energy index $s_h$ directly relates to the measured energy index by IceCube, $s_{\rm ob}$, which can take values $\in [2.19,3.09]$, at $1\sigma$ C.L., depending on the chosen data set. The energy break $\varepsilon_{\rm br}$, after correction for the redshift effect, is related to the threshold energy which varies from $\sim10$~TeV to $\sim100$~TeV among different data sets. Clearly, the most interesting neutrino data set will be the Cascade events (see the third row in Table~\ref{tab:nudata}) since its extension to low energies is experimentally confirmed, providing a proof that astrophysical neutrinos exist in this energy range. Finally, the normalization $A$, times a multiplicative constant in the cosmic evolution function $\mathcal{F}(z)$ parameterizing the percentage of population contributing to the diffuse neutrino flux, is related to the normalization of the observed diffuse flux by IceCube, i.e. $\Phi_{\rm astro}$ in Eq.~(\ref{eq:icflux}). We would like to emphasize that although the spectrum in Eq.~(\ref{eq:nuspec}) can be realized in the photohadronic ($p\gamma$) scenario for neutrino production by roughly assuming $\sim 6~{\rm keV}\,\left({\rm TeV}/\varepsilon_{\rm br}\right)$ for the energy of target photons, the employed spectrum is basically the {\it minimum} required to interpret the observed neutrino flux at IceCube. Note that a different choice for the behaviour above 10 PeV, provided it is consistent with null observations in this range, would not alter our conclusions. Similarly, the exact behaviour below $\varepsilon_{\rm br}$ is unknown: using a constant flux (rather than e.g. a different power law) below the threshold is the simplest choice but it is not essential, provided that the bulk of the energy carried by the flux is around $\varepsilon_{\rm br}$. 

Neutrino production through charged pion and kaon decay is associated to 
a $\gamma$-ray yield 
\begin{equation}\label{eq:nugamma}
\varepsilon_\gamma\, \frac{{\rm d}N_\gamma}{{\rm d}\varepsilon_\gamma} = \frac{4}{3K} \left[\varepsilon_\nu\, \frac{{\rm d}N_\nu}{{\rm d}\varepsilon_\nu}\right]_{\varepsilon_\nu = \varepsilon_\gamma/2}~,
\end{equation}
where the neutral to charged pion ratio $K\approx 1$ for the $p\gamma$ scenario. Note that Eq.~(\ref{eq:nugamma}) is a {\it minimal} Ansatz on the $\gamma$-ray flux, since it ignores any further leptonic contribution which would have no neutrino counterpart. While propagating from the sources (assumed transparent) to the Earth, $\gamma$-rays initiate electromagnetic cascades by pair-production on and inverse-Compton scattering off the CMB and EBL, resulting in a diffuse $\gamma$-ray flux at the Earth with energies $\lesssim1$~TeV. The exact spectral shape of the diffuse $\gamma$-ray flux depends on the cosmic evolution $\mathcal{F}(z)$ and the distance to the sources. In the limit of fully developed cascades, with some approximations one can derive analytically a universal spectral shape for the diffuse $\gamma$-ray flux~\cite{Berezinsky:1975zz,Ginzburg:1990sk,Berezinsky:2016inf}. For our quantitative purposes, we require a more precise spectral calculation, a task we tackle numerically via the public \textsc{$\gamma$-Cascade} code~\cite{Blanco:2019oa}. Note that in Eq.~(\ref{eq:nudiff}) there is only one unknown normalization at the RHS that has to be fitted to the data. Once we assume a shape for $\mathcal{F}(z)$, we normalize it via 
\begin{equation}\label{eq:norm}
\int {\rm d}z\, \frac{{\rm d}\mathcal{V}_c}{{\rm d}z}\, \mathcal{F}(z) =1\,,
\end{equation}
so that the remaining relevant normalization to be adjusted to the data is $A$ from ${\rm d}N_\nu/{\rm d}\varepsilon_\nu$ in Eq.~(\ref{eq:nuspec}). The normalization of Eq.~(\ref{eq:norm}) effectively introduces one `equivalent' source for the whole population, whose luminosity equates the total bolometric luminosity of the population and is given by
\begin{equation}\label{eq:luminosity}
   L_\nu = \int_{\varepsilon_{\rm min}}^{\varepsilon_{\rm max}} {\rm d}\varepsilon_\nu \; \varepsilon_\nu \; \frac{{\rm d}N_\nu}{{\rm d}\varepsilon_\nu}~.
\end{equation}

Note that the cascaded flux is an \textit{extra} contribution to the EGB, and should be added to the contribution of sources, such as blazars, which are already included in the current modelings. This is because the included conventional contributions to EGB are the ``guaranteed'' contributions inferred from $\gamma$-ray studies at TeV energies or below; they are present irrespective of, say, blazars being or not responsible for the observed neutrino flux by IceCube. However, if blazars happen to be the source of IceCube neutrinos, their emission spectrum should also extend from ${\cal O}$(10) TeV up to $\gtrsim 1$~PeV, thus inducing a further contribution to the EGB via the electromagnetic cascade process.
\section{\label{sec:rob}The bounds and their robustness}

\subsection{\label{sec:sfr}The SFR case}
The diffuse $\gamma$-ray flux from electromagnetic cascades contribute to the EGB measured by Fermi-LAT, adding up to other established \textit{conventional} contributions discussed in section~\ref{sec:EGBdata}. An upper limit on the cascade $\gamma$-ray diffuse flux is derived via a $\chi^2$ function analogous to Eq.~(\ref{eq:chi2-egb}), now containing also a cascade term $j={\rm cas}$, with $F^{\rm cas}=F^{\rm cas}(s_h,\varepsilon_{\rm br},A)$. The $2\sigma$ C.L. limit on two of these parameters (say $s_h$ and $A$, or its equivalent $\Phi_{\rm astro}$), by fixing the third parameter, can be derived by requiring $\Delta\chi^2=\chi^2-\chi^2_{\rm min} < 6.18$, where $\chi^2_{\rm min}$ is the minimum value of the $\chi^2$ scanning over all the cascade flux contributions.

Recently, constraints on the IceCube neutrino sources have been derived  in~\cite{Capanema:2020rjj}, using nearly the same argument we had developed and: i) The star forming rate (SFR)~\cite{Hopkins:2006bw,Yuksel:2008cu} has been used for the cosmic evolution $\mathcal{F}(z)$; ii) the whole energy range of EGB data (for foreground modeling \textbf{A}) has been considered; iii) the $\chi^2$ function used in~\cite{Capanema:2020rjj} further differs from the current form via the following: in~\cite{Capanema:2020rjj} a single multiplicative nuisance parameter $\alpha$ has been used {\it for the sum} of all the conventional contributions. Also, the conventional contributions in~\cite{Capanema:2020rjj} are based on Ref.~\cite{Ajello:2015mfa}, Ref.~\cite{Inoue:2011bm} and Ref.~\cite{Ackermann:2012vca}, respectively, for blazars, mAGNs and SFGs.

In this section we investigate the robustness of the limits derived in~\cite{Capanema:2020rjj}. In particular, besides updating the conventional contributions to EGB to more recent computations: i) We investigate the impact of various Galactic foreground modelings in the derivation of EGB data. ii) We allow for two independent nuisance parameters, which take into account the uncertainties in each contribution separately. iii) We discuss the dependence of the constraints on the considered energy range of EGB data in the analysis. iv) We comment on the independence of the obtained bounds from the actual astrophysical model adopted for the conventional blazar contribution to the EGB.

\begin{figure}[t!]
\centering
\subfloat{
\includegraphics[width=0.5\textwidth]{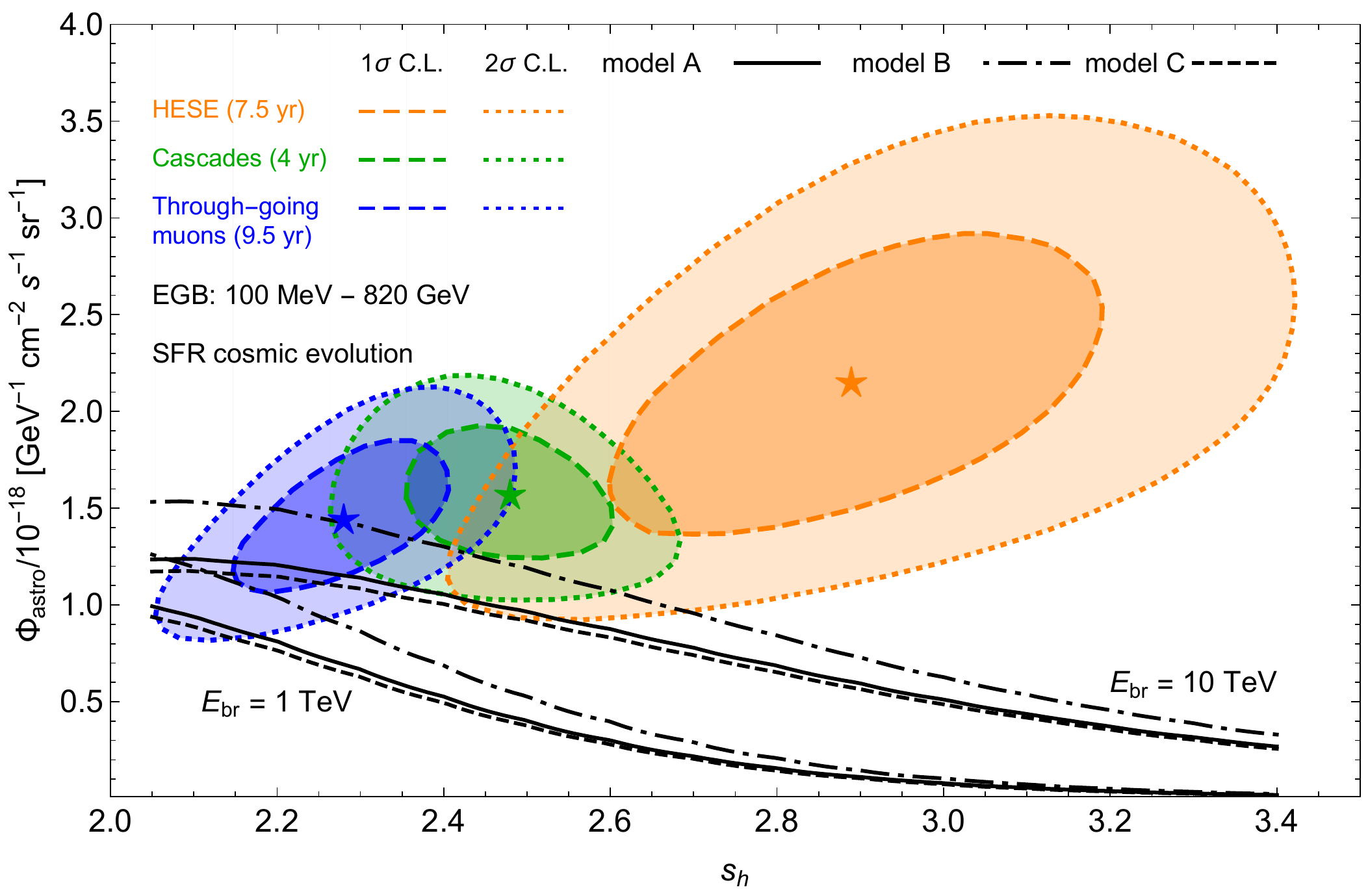}
\label{fig:FG-all}
}
\subfloat{
\includegraphics[width=0.5\textwidth]{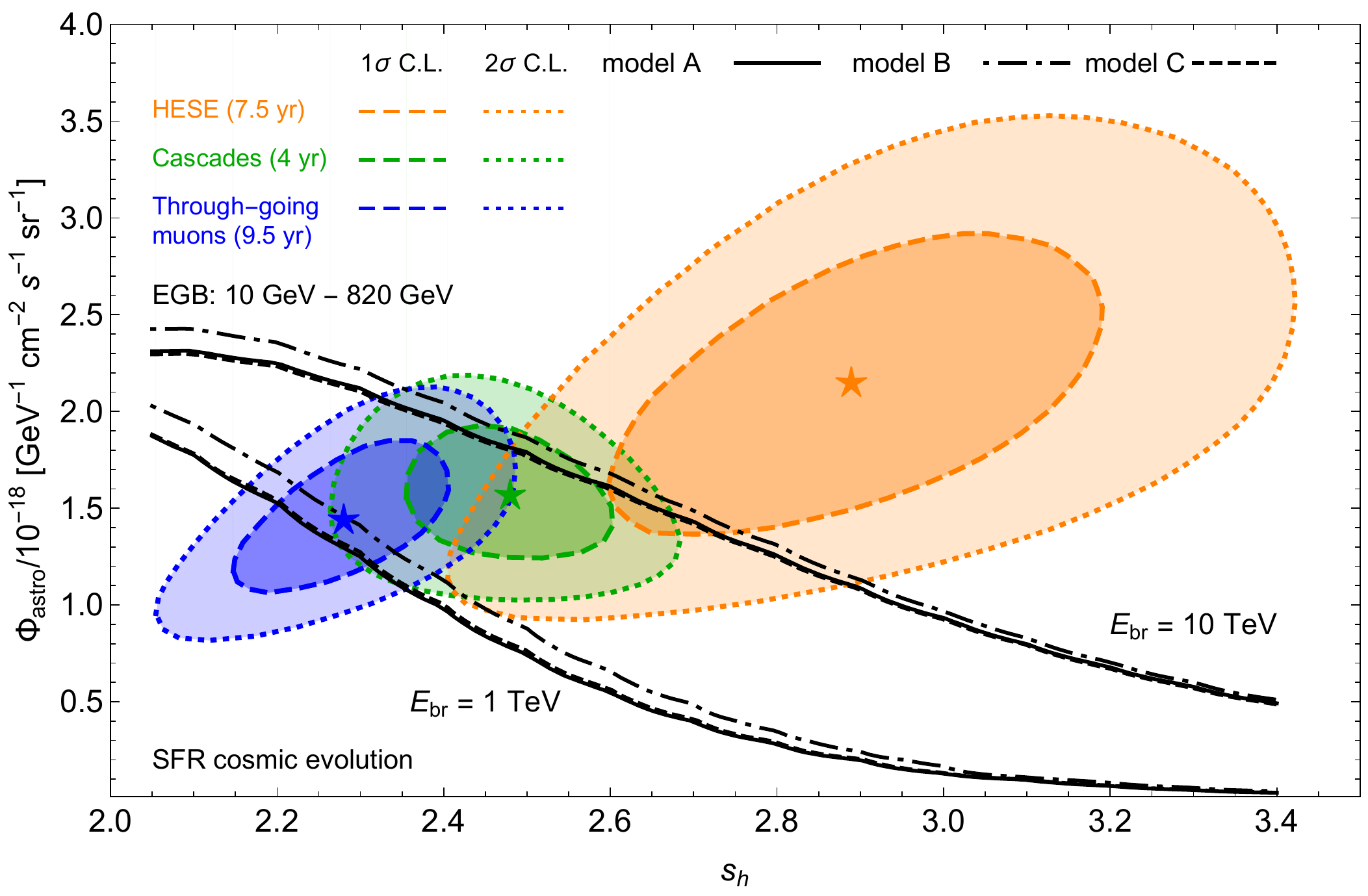}
\label{fig:FG-10}
}
\caption{\label{fig:FGmodels}Limits, at $2\sigma$ C.L., in the $(s_h,\Phi_{\rm astro})$  plane for SFR cosmic evolution of sources. The black curves show the limits for the three foreground modelings and for fixed threshold energies in neutrino spectrum $E_{\rm br}=1$~TeV and 10~TeV. The shaded regions show the $1\sigma$ and $2\sigma$ allowed regions of IceCube data sets. The left (right) panel is for the analysis of EGB data in the energy range 100 MeV - 820 GeV (10 GeV - 820 GeV).}
\end{figure}

\begin{figure}[t!]
\centering
\subfloat{
\includegraphics[width=0.65\textwidth]{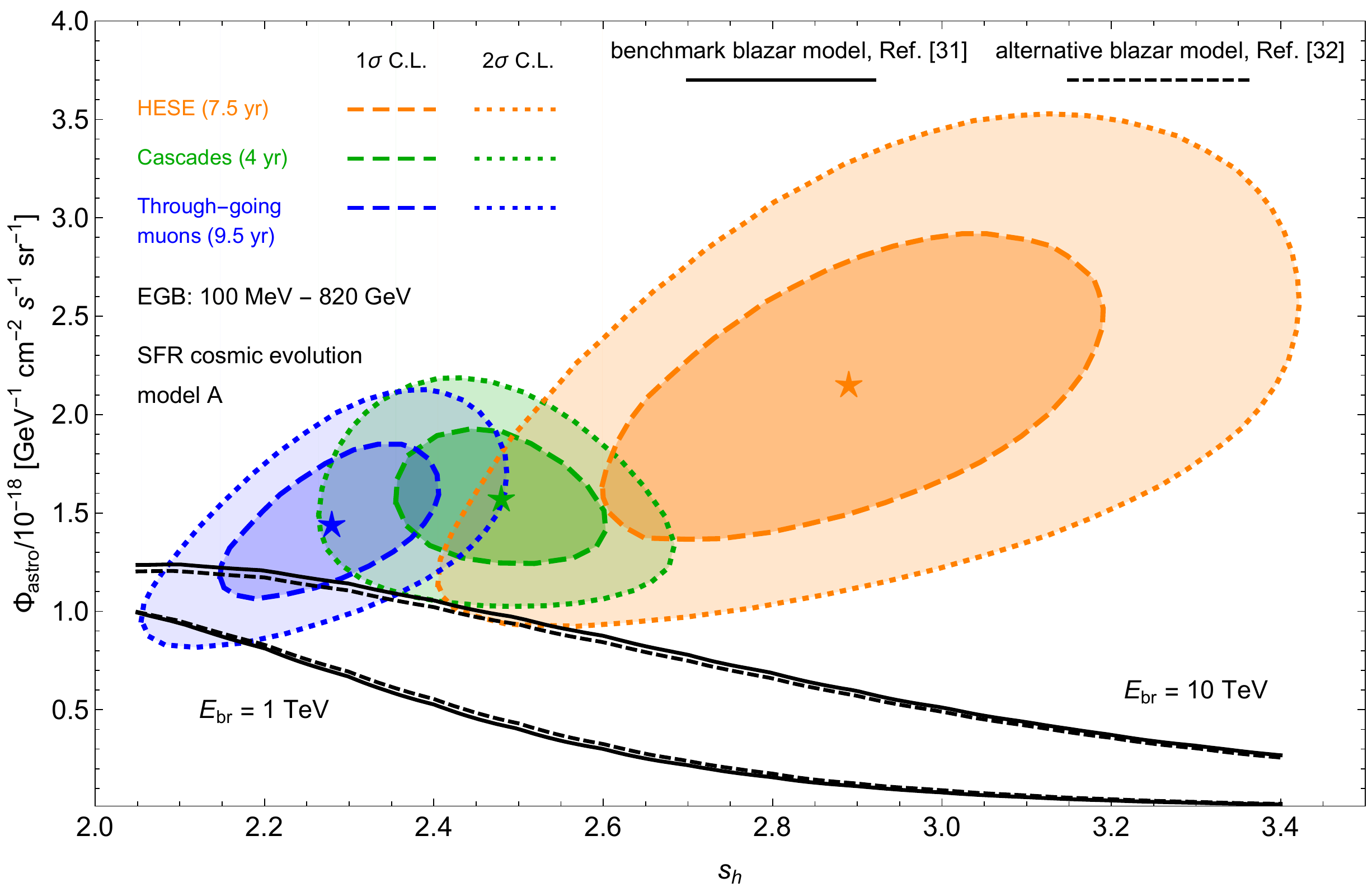}
\label{fig:FG-v2}
}
\caption{\label{fig:FGv2} As in the right panel of Figure~\ref{fig:FGmodels}, but comparing our benchmark blazar model from Ref.~\cite{Qu:2019zln} with the one reported in~\cite{DiMauro:2015tfa}.}
\end{figure}

The dependence of the constraints on the {\it foreground modeling} in the extraction of EGB data is shown in Figure~\ref{fig:FGmodels}, where the black curves depict the limits, at $95\%$ C.L., in the $(s_h,\Phi_{\rm astro})$ plane for fixed break energies in neutrino spectrum. The shaded regions show the allowed regions of IceCube data sets at $1\sigma$ and $2\sigma$ C.L. The left (right) panel is for the analysis of EGB data in the energy range 100 MeV - 820 GeV (10 GeV - 820 GeV). In this figure we assume the SFR cosmic evolution for the sources~\cite{Hopkins:2006bw,Yuksel:2008cu}. Since the SFR cosmic evolution peaks at $z\sim1-2$, the labelled $E_{\rm br}$ in Figure~\ref{fig:FGmodels} corresponds to the redshifted transformation of $\varepsilon_{\rm br}$ entering Eq.~(\ref{eq:nuspec}) (see~\cite{Capanema:2020rjj} for details). In both panels a log-normal pull-term in $\chi^2$ has been used, although the results are almost the same for a Gaussian type of pull-term. The following remarks are in order: i) The limits are almost independent of the foreground modeling. The foreground model \textbf{B} slightly relaxes the limits, especially if the whole energy EGB data is used, mostly as a consequence of the large error bars at low energies. ii) The solid lines of the left panel can be directly compared to the results of~\cite{Capanema:2020rjj}, where the same setup has been used, but with a single nuisance parameter for the conventional contributions to the EGB. As a result of the enhanced flexibility of the astrophysical background models, the limits are relaxed by $\sim40\%$ with respect to~\cite{Capanema:2020rjj}, and the tension for $E_{\rm br}=1$~TeV (10 TeV), while still present, now drops to $\gtrsim3\sigma$ ($\sim2\sigma$) at $s_h=2.53$ which is the best-fit value of the Cascade data set. For larger $s_h$ values, the tension aggravates, rising for instance to $\gtrsim4\sigma$ for  $s_h=2.9$, which is the HESE best-fit value. iii) The constraints derived from the 10 GeV - 820 GeV energy range of EGB, shown in the right panel of Figure~\ref{fig:FGmodels}, are weaker than the constraints derived from the whole EGB energy range. The $E_{\rm br}=10$~TeV is now compatible with the Cascade data set while the tension reduces to $\sim2\sigma$ for $E_{\rm br}=1$~TeV. We have also checked that by limiting the analysis to the 10 GeV - 820 GeV range and considering only the blazars contribution, the bounds would be similar to the right panel of Figure~\ref{fig:FGmodels}.
iv) Had we adopted a different model for the conventional blazar contribution to the EGB, it would not have appreciably changed our conclusions. This is illustrated in Figure~\ref{fig:FGv2}, where we compare our benchmark results with the ones one would obtain if using the blazar modeling reported in~\cite{DiMauro:2015tfa}. This insensitivity is largely due to the similar spectral shape and roughly consistent normalization, within errors, that all modern viable blazar models manifest. Significant differences are only found at low energy, where anyway blazars are not the dominant contribution to the EGB and thus have a negligible influence on the bounds.

From the results of this section we can conclude that, by fitting the EGB in its entire energy range with state-of-the-art models for the conventional contributions, the tension between the EGB and IceCube data persists at at least $(2-3)\sigma$ for $E_{\rm br}\sim(1-10)$~TeV, assuming that the  neutrinos originate from transparent astrophysical sources with a $z$-distribution resembling the SFR cosmic evolution. By restricting oneself to $>10$~GeV EGB data, the tension can be alleviated, but remains substantial at least in the case of $E_{\rm br}\sim 1$~TeV. Loosening the bounds this way requires however to pay a price: the low energy ($<10$~GeV) EGB data cannot be interpreted by (current models of) the mAGN and SFG contributions. Either current models of these classes of sources are inadequate, or one or more new classes of sources has a significant contribution at low energies.

However, the conventional contributions considered in this paper are among the most conservative evaluations in the literature. Had we used larger available estimates, the limits reported in this paper would have been stronger. For example, the mAGN contribution that we use, based on~\cite{DiMauro:2013xta}, is smaller than the one calculated in~\cite{Hooper:2016gjy}. The same applies to SFG and blazar contributions (see e.g.~\cite{Cholis:2013ena} for an analysis with larger contributions). Also, there are at least two more ``guaranteed'' contributions that we conservatively neglected: a) The leptonic counterpart to the hadronic flux responsible for the neutrino production, discussed after Eq.~(\ref{eq:nugamma}). b) The $\gamma$-ray production in the propagation of ultra high energy cosmic rays (UHECRs), see~\cite{Ahlers:2011sd}, which strongly depends on the composition of UHECRs. In the light of these considerations, it appears likely that the tension between the two data sets, EGB and IceCube, is real and it is worth exploring alternative solutions to it.

\subsection{\label{sec:bl}The case of BL Lacs}

\begin{figure}[t!]
\centering
\includegraphics[width=0.6\textwidth]{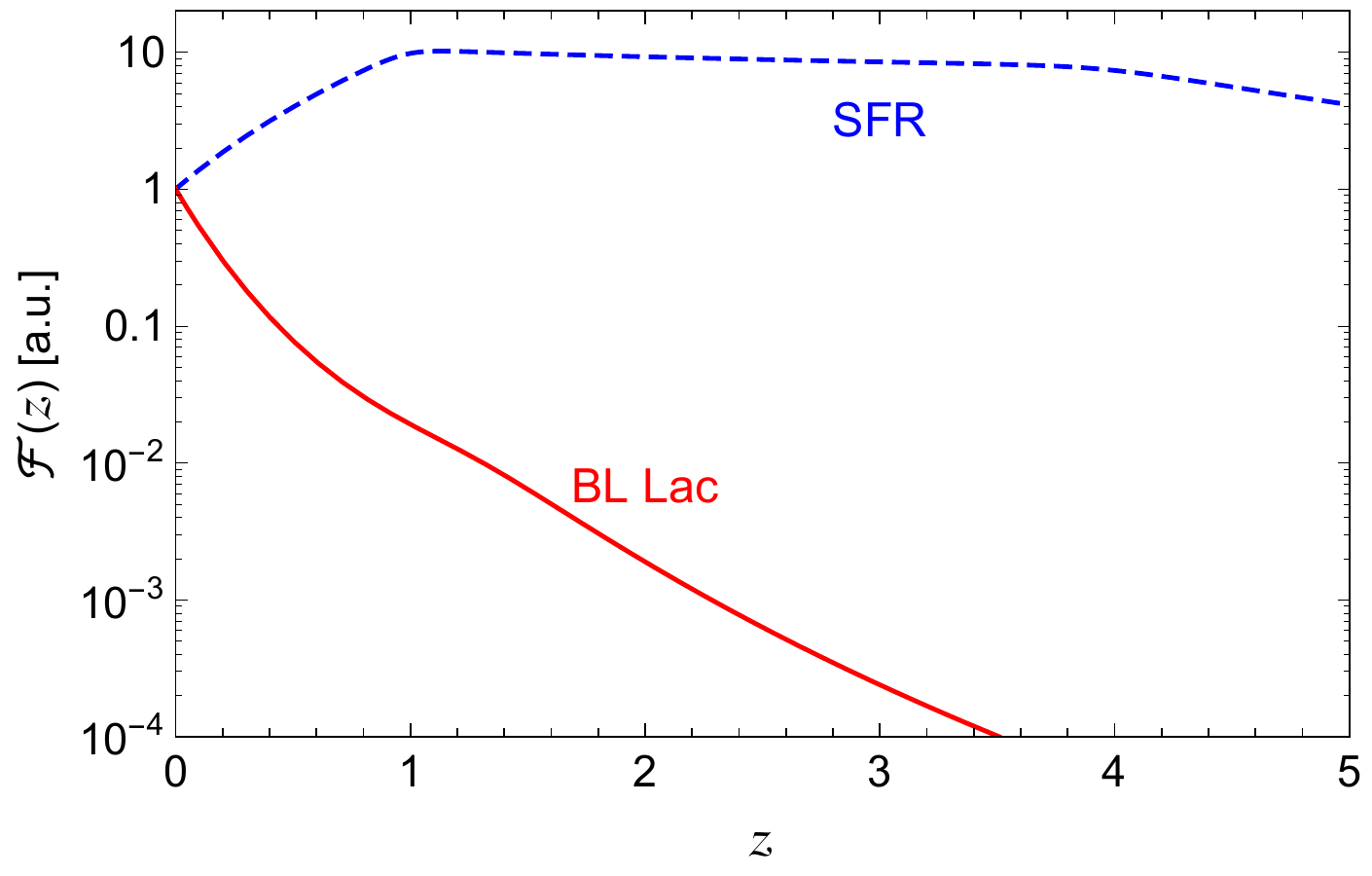}
\caption{\label{fig:fz}The SFR (blue-dashed)~\cite{Hopkins:2006bw,Yuksel:2008cu} and BL Lac (red-solid)~\cite{Ajello:2013lka} cosmic evolution functions. Both functions are normalized such that $\mathcal{F}(z=0)=1$ a.u.}
\end{figure}

From the robustness analysis performed in the previous section, one may naively anticipate that the tension can be lifted if most of the sources of neutrinos are located much closer. In the SFR cosmic evolution, most of the sources are located at cosmological distances, $z\sim 1-2$, which results in the full development of electromagnetic cascade with significant contribution at $\sim100$~GeV. Supposedly, assuming a cosmic evolution featuring a ``local universe'' population of  sources would lessen the cascade flux, thus allowing the EGB and low energy IceCube data to be reconciled. In order to understand if known and relatively closer sources are enough to lift the constraints, in this section we take the BL Lac cosmic evolution~\cite{Ajello:2013lka} that peaks at $z\simeq0$. Figure~\ref{fig:fz} shows the SFR (blue-dashed) and BL Lac (red-solid) cosmic evolution functions where both the cases are normalized at $z=0$ to unity. More specifically, our red-solid curve represents the luminosity function provided in~\cite{Ajello:2013lka}, integrated over all possible spectral indices and over the entire luminosity interval (from  $7 \times 10^{43}$ erg/s to $10^{52}$ erg/s). Note that this implies an extrapolation of the distribution of low-luminosity BL Lacs ($L_\gamma \lesssim 10^{45}$ erg/s) beyond $z \simeq 0.5$, where Fermi loses sensitivity. The extrapolation is such that, at redshifts higher than the peak redshift of each luminosity interval, all luminosity bins decline essentially with the same slope.  When integrating the luminosity function over low luminosities only, we find a negative redshift evolution very close to the red-solid curve in Figure~\ref{fig:fz}, as predicted by the model in~\cite{Ajello:2013lka}. Otherwise said, the bolometric evolution is dominated by the evolution of the low-luminosity BL Lacs.
Although there is no direct evidence for the correctness of this specific model at high-$z$, the above qualitative conclusion is likely robust, since to alter it one would require a significant (and puzzling) growth of the high-luminosity bins at high-$z$, for which no indication is available. For our purpose, we content ourselves to use this as a plausible, if not fully correct, benchmark of the BL Lac source class peaking at $z=0$ and monotonically decreasing afterwards.   

\begin{figure}[t!]
\centering
\subfloat{
\includegraphics[width=0.5\textwidth]{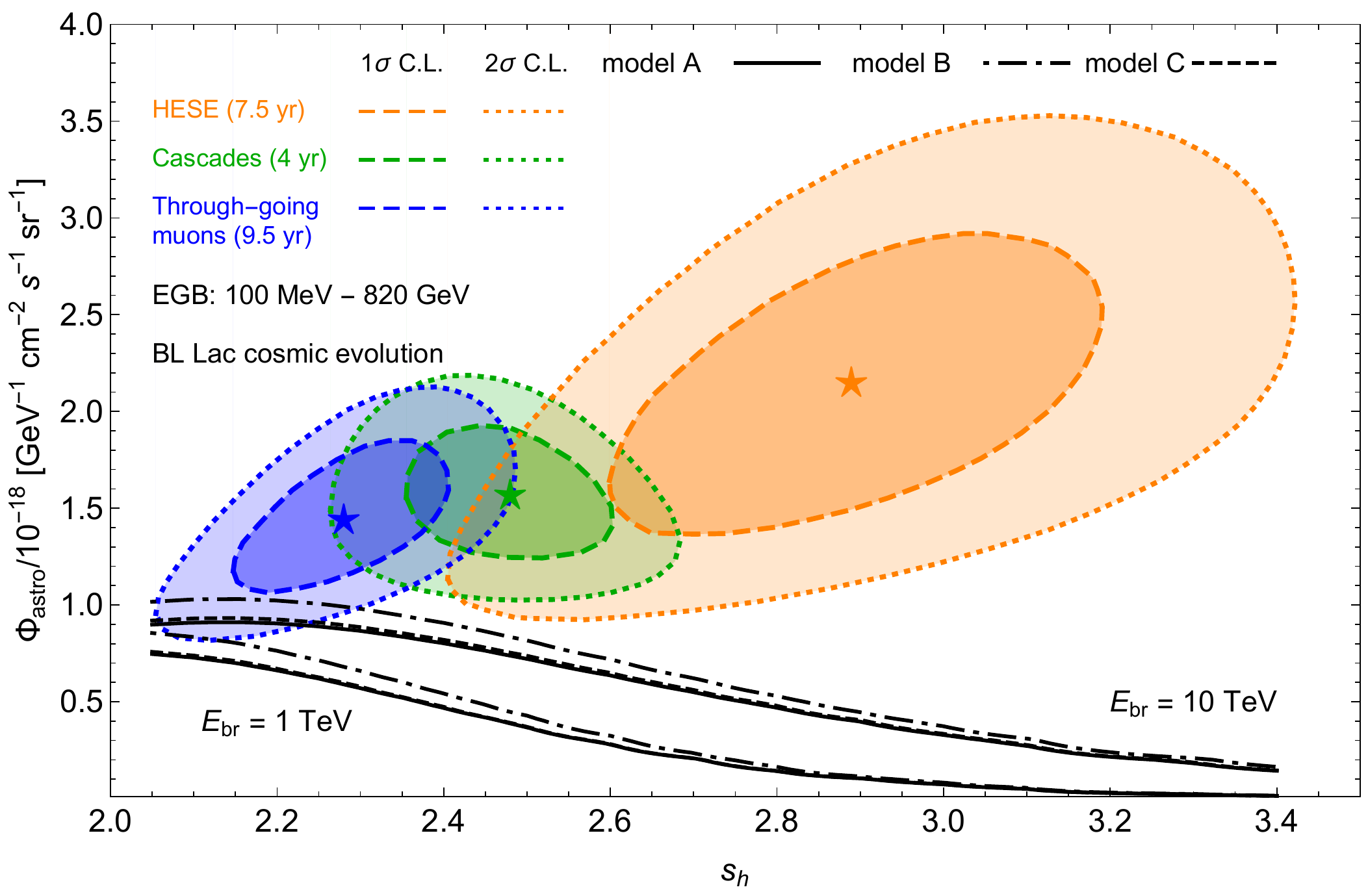}
}
\subfloat{
\includegraphics[width=0.5\textwidth]{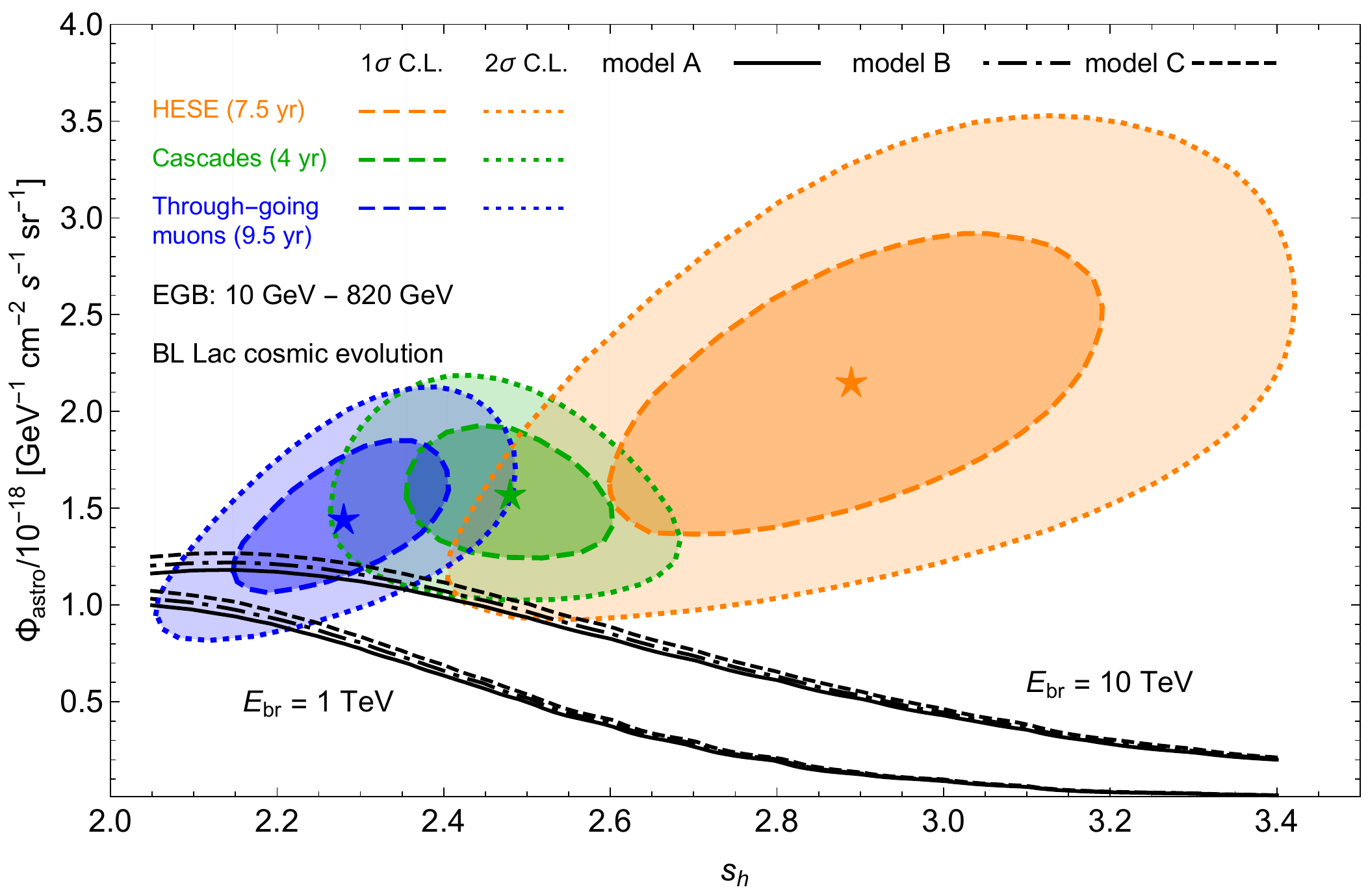}
}
\caption{\label{fig:FGmodels-BL}The same as Figure~\ref{fig:FGmodels}, but for BL Lac cosmic evolution of sources.} 
\end{figure}

\begin{figure}[t!]
\centering
\includegraphics[width=0.8\textwidth]{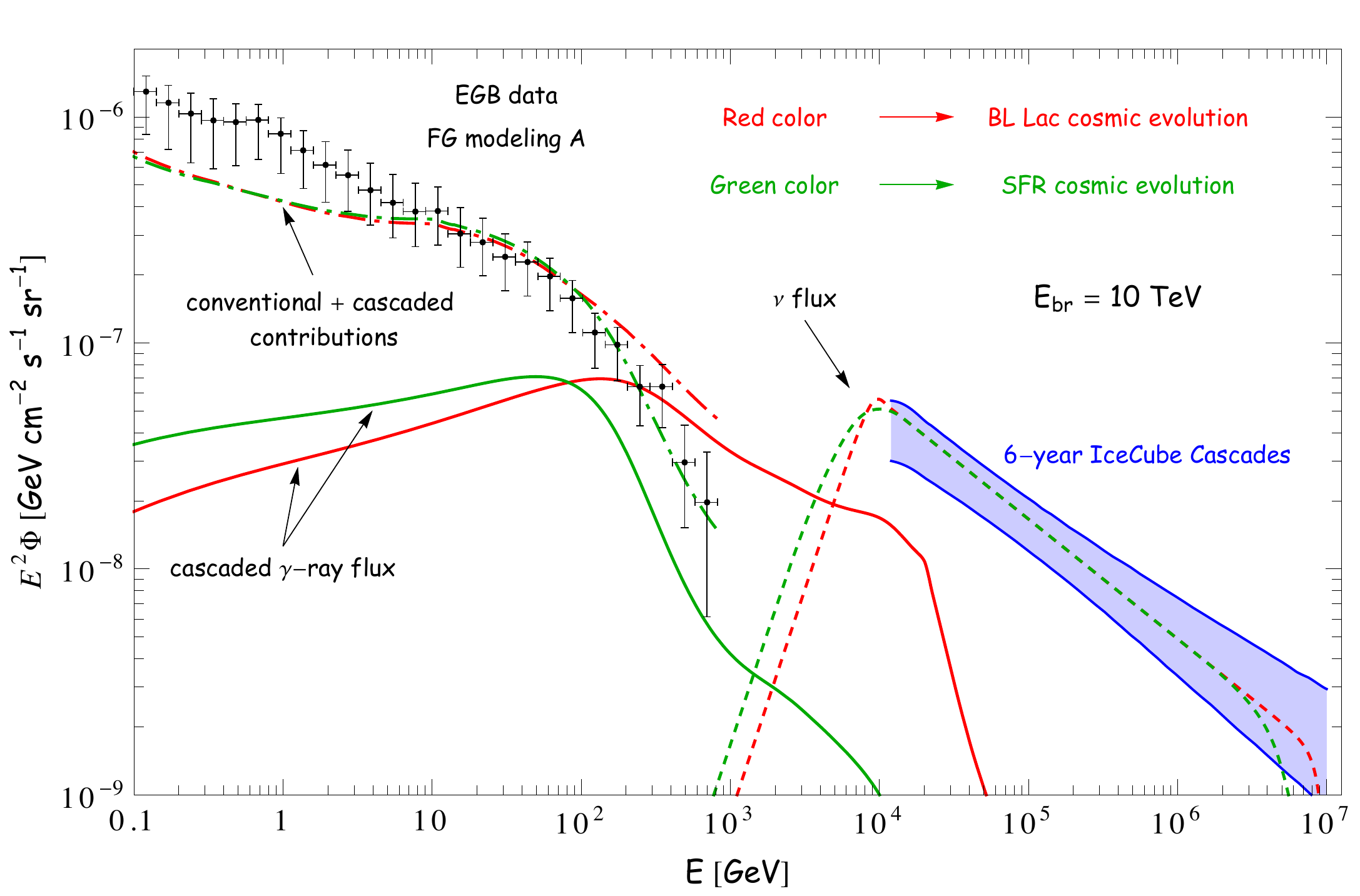}
\caption{\label{fig:spec10}Spectra of neutrino, cascade $\gamma$-ray and total (cascade + conventional) $\gamma$-ray fluxes depicted respectively by the dashed, solid and dot-dashed curves. The energy break is fixed to $E_{\rm br}=10$~TeV, while $s_h$ and $\Phi_{\rm astro}$ are fixed to their best-fit values for the Cascade data set of IceCube. Data points are EGB data from Fermi-LAT for foreground modeling \textbf{A}. The red and green colors correspond to BL Lac and SFR cosmic evolutions, respectively. The blue shaded region depicts the $1\sigma$ C.L. uncertainty from the 6-year Cascade data set of IceCube.}
\end{figure}

Figure~\ref{fig:FGmodels-BL} shows the limits ($2\sigma$ C.L.) in the $(s_h,\Phi_{\rm astro})$ plane for BL Lac cosmic evolution. For BL Lacs, the redshift effect is negligible and $E_{\rm br}\simeq \varepsilon_{\rm br}$. The dependence on foreground modeling is less than for the SFR evolution. Also, the limits based on the whole energy range of EGB and the $>10$~GeV range are similar, mainly since the constraints originate from the high energy part of the EGB data in the BL Lac case. The cascade $\gamma$-ray fluxes from SFR and BL Lac cosmic evolutions are qualitatively different. Figure~\ref{fig:spec10} illustrates the cascade fluxes for $E_{\rm br}=10$~TeV, with $s_h$ and $\Phi_{\rm astro}$ fixed to their best fit values in the Cascade data set from IceCube. The shaded blue region shows their $1\sigma$ C.L. uncertainty. The neutrino, cascade $\gamma$-ray and total $\gamma$-ray (conventional + cascade) fluxes are shown, respectively, by the dashed, solid and dot-dashed curves. The red and green colors correspond to, respectively, BL Lac and SFR cosmic evolutions. It can be seen from Figure~\ref{fig:spec10} that the main contribution to EGB data for the SFR evolution is at $\sim100$~GeV and drops soon thereafter, while for the BL Lac evolution, the higher energy ($\gtrsim100$~GeV) contribution is more pronounced. For the BL Lac evolution, despite the cascade being only partly developed as a consequence of the peak in the BL Lac distribution at $z\simeq0$, it falls where the highest energy data of the EGB are particularly constraining. 

Another manifestation of the tension between EGB and the models fitting the IceCube flux can be seen in the low-energy part of Figure~\ref{fig:spec10}, with the dot-dashed curves undershooting the EGB data, even though the whole energy range of EGB has been considered in their derivation. The reason is that the normalization nuisance parameters of blazar and mAGN contributions are pushed down in order to make some room for the cascade flux at high energies, at the expense of the description of the low-energy EGB data. 

These results suggest two possible ways through which one may hope to ease the tension: i) To reduce the relative weight of the cascade with respect to the EGB via sources at high-$z$, by effectively ``moving the green curve'' in Figure~\ref{fig:spec10} to the left; ii) Alternatively, to reduce the development of the cascade by moving the sources closer, thus effectively ``moving the red curve'' in Figure~\ref{fig:spec10} to the right.
Next, we explore these two possibilities, discussing their plausibility. 

\section{\label{sec:highz}High-$z$ sources}

\begin{figure}[t!]
\centering
\includegraphics[width=0.8\textwidth]{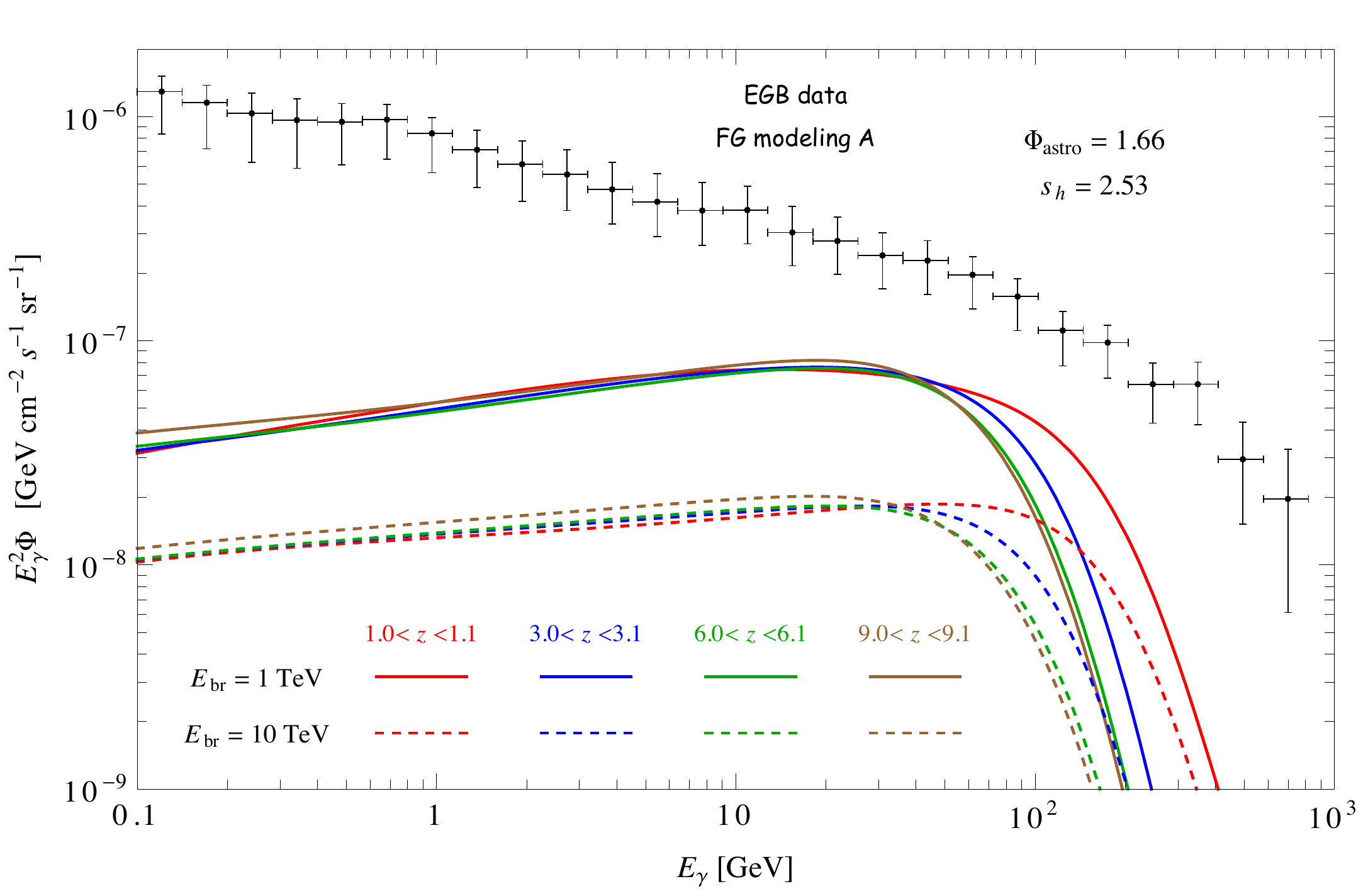}
\caption{\label{fig:hz}The cascade flux for a population of sources with flat distribution in $z_{\rm min} < z < z_{\rm max}$. The solid and dashed curves correspond to $E_{\rm br}=1$~TeV and $E_{\rm br}=10$~TeV, respectively. For all the curves, $\Phi_{\rm astro}=1.66$ and $s_h=2.53$ which correspond to the best-fit values of IceCube's Cascade data set.}
\end{figure}

For distant sources located at high redshifts the cascade process fully develops. Compared to the SFR case and even more to the BL Lac case (in Fig.~\ref{fig:spec10}), the diffuse $\gamma$-ray flux moves to lower energies, suppressing the emission at $E\gtrsim 100$~GeV, see Figure~\ref{fig:hz}. In fact, for a population of sources with redshifts $\gtrsim 3$, the spectrum of cascade photons approaches the universal form, calculated analytically in~\cite{Berezinsky:1975zz,Ginzburg:1990sk,Berezinsky:2016inf}, with $\Phi\propto E_\gamma^{-1.8}$ for $E_\gamma\lesssim30$~GeV and a cutoff at $\sim30$~GeV originating from the pair-production on EBL and CMB target photons, which have higher energies at large redshifts~\cite{Berezinsky:2016inf}. 

A preliminary remark is in order: sources at high-$z$ are automatically invisible in the high-energy gamma ray band, even if the source itself is transparent and $\gamma$-rays can escape from it. In fact, the evolution of CMB with redshift improves the pair-production efficiency in two ways: i) The CMB photon number density scales $\propto (1+z)^3$ leading to a smaller mean free path for high energy $\gamma$-rays. ii) The energy, or equivalently the temperature, of CMB photons increases $\propto (1+z)$, which lowers the threshold energy for pair-production and makes lower energy photons vulnerable to absorption. At $z\simeq0$ the mean free path for pair-production on CMB is quite small, $\sim4.8$~kpc and the threshold energy is $\simeq400$~TeV, where for both values we assumed the energy and cross section at the peak of the CMB spectrum. At higher redshifts, the lower threshold energy and higher number density rapidly makes the universe opaque to $\gamma$-rays, such that $\gamma$-rays with energy $\gtrsim1$~TeV will be eventually absorbed merely by pair-production on CMB~\cite{Ruffini:2015oha,Tizchang:2017oky}. On the top of that, at $z\lesssim4$ the interaction with the EBL is also relevant, notably up to $\sim 10$~TeV, strengthening the absorption. In summary, independent of whether the high-$z$ distributed sources can lessen the tension between IceCube data and the EGB or not, the absorption on CMB and EBL renders these sources invisible to us at $\gtrsim10$~GeV (see~\cite{Ruffini:2015oha}). This will make any attempt to find directional correlation of IceCube neutrinos with these sources pointless, unless considering observations in the MeV band or below, which requires strongly model-dependent assumptions.

In order to visualize the characteristics of the cascade flux from high-$z$ sources, we use a toy-model where all the sources are located in a range of redshift $z_{\rm min}<z<z_{\rm max}$, with a flat distribution ($\mathcal{F}(z)=$~constant) in this range. Figure~\ref{fig:hz} shows the cascaded fluxes in these models, where the neutrino spectrum is fixed to the best-fit of IceCube's Cascade data set $\Phi_{\rm astro}=1.66$ and $s_h=2.53$. The solid (dashed) curves correspond to $E_{\rm br}=1$~TeV (10~TeV). It is clear that the cascaded flux saturates to its universal form for $z\gtrsim3$. The compatibility of these cascaded fluxes with the EGB data can be quantified by performing a $\chi^2$ analysis similar to section~\ref{sec:rob}. For $E_{\rm br}=10$~TeV, the cascaded flux from sources with $z\gtrsim3$ matching the best-fit point of IceCube's Cascade data set are indeed compatible with the EGB data. In both the energy ranges of EGB data (above 100 MeV or above 10 GeV as discussed in section~\ref{sec:rob}), the cascaded flux contribution worsens the fit by only $\Delta\chi^2\sim3.5$, which is statistically insignificant. This conclusion remains valid for any $\mathcal{F}(z)$ of sources restricted to $z\gtrsim3$, since the shape of cascaded flux is close to its universal form for all large $z$.  

However, the case of $E_{\rm br}=1$~TeV is quite different: in this case, the cascaded flux for $\Phi_{\rm astro}=1.66$ and $s_h=2.53$ worsens the fit to EGB data by $\Delta\chi^2\sim40$, indicative of a strong tension. The worsening of the fit for $E_{\rm br}=1$~TeV is independent of which energy range of EGB data is included in the analysis. This strong tension can be qualitatively understood as follows: the cascaded flux for high-$z$ sources effectively contributes only to 3-4 bins of EGB data (around the peak at $\sim30$~GeV) which statistically can be easily constrained, if the normalization is high enough. This is contrary to the SFR case, where the cascaded flux mimics the EGB spectral shape $\gtrsim30$~GeV and thus can be accommodated in the fit by a small change in the normalization of the blazar contribution, within its band of uncertainty. We thus conclude that for $E_{\rm br}=1$~TeV, distant sources with $z\gtrsim3$ have the same level of tension with EGB data as the BL Lac and SFR distributed sources (see Figures~\ref{fig:FGmodels} and \ref{fig:FGmodels-BL}).

Although at face value high redshift sources can reconcile the IceCube and EGB data for $E_{\rm br}=10$~TeV, this comes at the price of an increased energy budget for this population. The required total neutrino luminosity of the population can be derived from IceCube observations for a fixed $\Phi_{\rm astro}$, $s_h$ and $E_{\rm br}$. Clearly, the higher the $z$, the larger the required bolometric neutrino luminosity. For example, for a flat distribution in $z_{\rm min}<z<z_{\rm max}$, the required total neutrino luminosity to interpret the best-fit of IceCube's Cascade data set (with $E_{\rm br}=10$~TeV) varies from $2.9\times10^{49}$ erg/s for $1<z<1.1$ to $5.9\times10^{51}$ erg/s for $9.9<z<10$. The required luminosity for $E_{\rm br}=1$~TeV is a factor of $\sim4$ larger. The relation between the neutrino luminosity, $L_\nu$, and the all-particle luminosity, $L_{\rm tot}$, of a source is generally highly model-dependent: For instance, for stellar objects it varies from $L_{\nu}\simeq99\% L_{\rm tot}$ for core-collapse supernovae to $L_{\nu}\simeq0.1\% L_{\rm tot}$ for main-sequence hydrogen burning stars such as Sun. 

To get an idea of how large these luminosity requirements are, let us compare with a physically motivated case. The most luminous sources known to exist at high redshifts are associated to supermassive black holes (SMBH). The mass function of SMBH (intended here as having masses above $\sim 10^6\,M_\odot$, where $M_\odot$ is the solar mass), at $z=6$, can be described as
\begin{equation}
\frac{{\rm d} n_{\rm BH}}{{\rm d} \log_{10} m}=\kappa\, m^\alpha e^{-m}\,,\label{paramet}
\end{equation}
where $\kappa=1.23\times 10^{-8}\,{\rm Mpc}^{-3}$, $\alpha=-1.03$ and $m\equiv M/M_*$, with $M_*=2.24\times 10^{9}\,M_\odot$ (see Ref.~\cite{2010AJ....140..546W} or equivalently Figure~2 in Ref.~\cite{2011MNRAS.417.2085V}). If to each SMBH of mass $m$ is associated a luminosity $L(m)$, the total luminosity $L_{\bullet,{\rm tot}}$ associated to this population at $z_\bullet=6$ is
\begin{equation}\label{eq:smbhpoplum}
L_{\bullet,{\rm tot}}=\iint {\rm d}m\, {\rm d}z\, \frac{{\rm d}\mathcal{V}_c}{{\rm d}z} \frac{{\rm d} n_{\rm BH}}{{\rm d}  m}\, L(m)= \frac{4\pi \left[d_c(z_\bullet)\right]^2}{E(z_\bullet)} \left(\frac{c}{H_0}\right)\int {\rm d}m \, \frac{{\rm d} n_{\rm BH}}{{\rm d}  m} L(m)\,,
\end{equation}
where $H_0$ is the present-day Hubble expansion rate, $E(z)=\sqrt{\Omega_\Lambda+\Omega_m (1+z)^3}$ and $d_c(z)=(c/H_0)\int_0^z {\rm d}z/E(z)$, with $\Omega_\Lambda=0.7$ and $\Omega_m=0.3$. In the last step of Eq.~(\ref{eq:smbhpoplum}) we assumed `instantaneous' emission at $z_\bullet=6$. Although the bolometric luminosity $L(m)$ is unknown, let us consider two benchmarks:

\begin{itemize}
    \item It has been speculated that the birth of SMBH is via a `supermassive SN/GRB explosion' releasing an energy $\sim 10^{55}$~erg~\cite{Woods:2018lty}. Assuming such energy release only once in the universe's lifetime (and all the energy released instantaneously at $z=6$) leads to a constant $L(m)=2.3\times10^{37}$~erg/s, where from Eq.~(\ref{eq:smbhpoplum}) results to $L_{\bullet,{\rm tot}} = 1.8 \times 10^{44}$~erg/s. This luminosity is seven orders of magnitude away from satisfying the IceCube required luminosity for sources at $6<z<6.1$ and $E_{\rm br}=10$~TeV, which is $\sim 2\times 10^{51}$~erg/s.
    \item Another characteristic luminosity for a BH is the Eddington luminosity $L(M)=1.26\times 10^{38} M/M_\odot$ erg/s, which sets the scale at which spherically accreting material is balanced by electromagnetic pressure. This is believed to represent an upper limit to the luminosity via accretion, the most efficient energy conversion process known in astrophysics that can be sustained in a steady-state way. In this case, $L_{\bullet,{\rm tot}}=4.4\times10^{51}$~erg/s, implying that a $50\%$ channeling of the all-particle luminosity into neutrinos would be needed to fulfill IceCube's requirement for $E_{\rm br}=10$~TeV. Keeping in mind that photons must carry a similar energy as neutrinos, and that their parent protons/nuclei must necessarily carry more energy, this estimate suggests an unrealistically large energy budget required. Another element is that (a possibly dominant) part of the accretion energy is dissipated via the relatively low energy emission of the disk, rather than the kinetic energy of the jet, worsening the requirements.  
\end{itemize}

To summarize this section, a population of sources at high redshifts $z\gtrsim3$ could make the IceCube and EGB data compatible for $E_{\rm br}=10$~TeV, but not for $E_{\rm br}=1$~TeV. However, energy considerations suggest that finding an appropriate energy budget is extremely challenging, to say the least. Our energy budget argument is consistent with~\cite{Xiao:2016rvd} which concludes that a dominant contribution to the IceCube flux is possible only if sources as frequent as Pop-III hypernovae with explosion energy $>10^{53}$~erg are present at large redshifts, while our refined analysis of EGB data excludes even this scenario for $E_{\rm br}=1$~TeV. Similarly, the results of this section conform to the conclusion of~\cite{Chang:2016ljk} where IceCube neutrinos with energy $>25$~TeV have been attributed to distant sources.

\section{\label{sec:lowz}Low-$z$ sources}

The alternative possible route to ease the tension between the IceCube and EGB data is to assume that most of the sources are located in the local universe, at $z\simeq0$. This assumption effectively flattens the peak of cascaded flux present in SFR and high-$z$ distributed sources (and to some extent for BL Lac distribution). For low-$z$ sources, $\gamma$-rays with energy below a few hundreds of TeV suffer pair-production just on EBL, which is significant only for source distances above ${\cal O}$(10) Mpc. Provided that $E_{\rm br}\gtrsim 1\,$TeV, we expect thus that sources within $z\sim {\cal O}(10^{-3})$ do not lead to conflicts with the EGB. 

\begin{figure}[t!]
\centering
\includegraphics[width=0.8\textwidth]{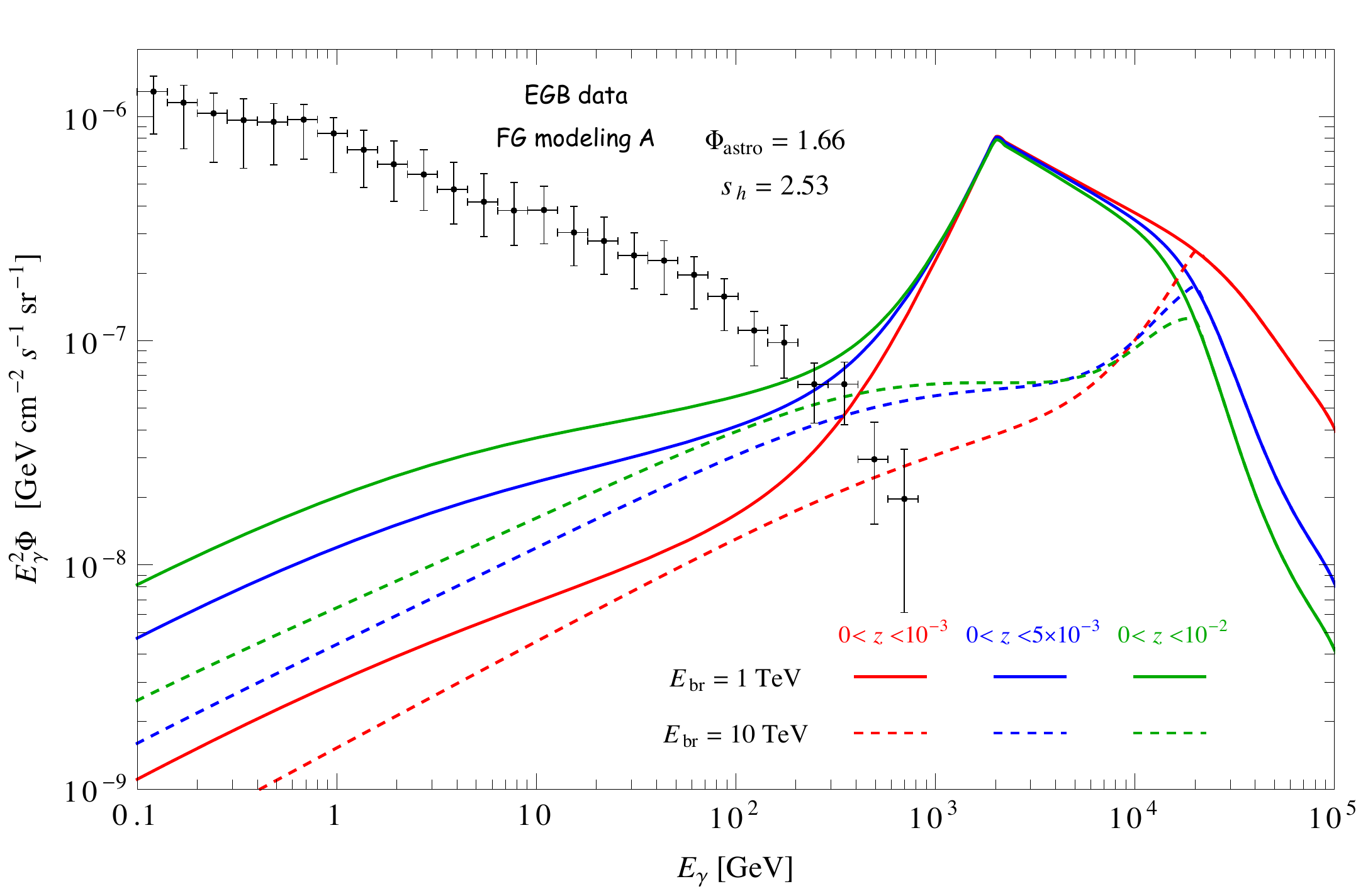}
\caption{\label{fig:zmax}The cascaded flux for a population of sources with flat distribution between $0 < z < z_{\rm max}$. The solid and dashed curves correspond to $E_{\rm br}=1$~TeV and $E_{\rm br}=10$~TeV, respectively. For all the curves, $\Phi_{\rm astro}=1.66$ and $s_h=2.53$, corresponding to the best-fit values of IceCube's Cascade data set.}
\end{figure}

At the same time, however, requiring a larger and larger fraction of sources which do not experience sizable opacity can lead to conflict with direct TeV-gamma-ray observations. First, let us quantify more precisely up to which distance the sources should be located in order to suppress the cascaded flux sufficiently to ease the tension with EGB. For this purpose, we assume that the sources are located in $0\leq z \leq z_{\rm max}$ with a flat distribution. This is reasonable since we expect $z_{\rm max}\sim 10^{-3} \ll 1$. Figure~\ref{fig:zmax} shows the cascaded flux for various $z_{\rm max}$ values, for $E_{\rm br}=1$~TeV (solid curves) and 10 TeV (dashed curves). For all the curves we assumed the best-fit point of IceCube's Cascade data set. The solid (dashed) curves have a break at 2 TeV (20 TeV) originating from $\varepsilon_\nu=\varepsilon_\gamma/2$ (see Eq.~(\ref{eq:nugamma})). From Figure~\ref{fig:zmax}, $E_{\rm br}=10$~TeV and $z_{\rm max}\leq 10^{-3}$ is found to be the viable region: despite saturating the average flux measured in the last two bins of EGB data this only leads to $\Delta\chi^2=4.1$ (for $z_{\rm max}= 10^{-3}$) due to the large measurement errors. On the other hand, by decreasing the energy break or increasing $z_{\rm max}$ the tension rapidly grows such that $(E_{\rm br},z_{\rm max})=(1~{\rm TeV},10^{-3})$ leads to $\Delta\chi^2\simeq100$ and $(E_{\rm br},z_{\rm max})=(10~{\rm TeV},5\times10^{-3})$ corresponds to $\Delta\chi^2\simeq27$. 

At face value, sources distributed within the `very local' universe, up to redshift $z\sim10^{-3}$ or equivalently up to a distance of $\sim4$~Mpc, can generate the observed neutrino flux by IceCube down to $\sim10$~TeV without creating a sizeable cascaded flux leading to tension with EGB data, but not down to $\sim1$~TeV. However, this scenario is untenable for other arguments: i) The distance $\lesssim4$~Mpc implies sources within the local group of Galaxies. It is hard to conceive why a population of extragalactic sources would vanish beyond a few Mpc, which is a snapshot over cosmological timescales typically involved in evolutive processes (like star formation, BH accreting material starvation, etc). ii) The red dashed curve in Figure~\ref{fig:zmax} is escaping exclusion due to the lack of EGB data above 820 GeV. The Fermi-LAT is however sensitive to multi-TeV $\gamma$-rays; in principle, it is possible to extend the EGB observations/bounds to these energies, although Monte Carlo simulations of the detector and reliable rejection of misidentified cosmic ray events in multi-TeV range are computationally quite expensive. Recently, it has been claimed in~\cite{Neronov:2018ibl,Neronov:2019ncc}, by analyzing the Fermi data, that the diffuse $\gamma$-ray flux extends to the multi-TeV range with practically the same intensity above 820~GeV; this has been interpreted as the Galactic counterpart of the IceCube neutrinos, although this appears in tension with the limits from KASCADE~\cite{Schatz:2003aw} and HAWC~\cite{Harding:2019tez}. Independent of its origin, by considering the claimed flux as credible determination in the multi-TeV range, it would exclude even the $(E_{\rm br},z_{\rm max})=(10~{\rm TeV},10^{-3})$ case. iii) The product of the density of sources in the local universe, $\rho_0$, and neutrino luminosity of sources, $L_\nu$, is fixed by the IceCube observation to $\rho_0 \left(E_\nu {\rm d}L_\nu/{\rm d}E_\nu\right)\simeq 8\times10^{35}$~erg/s/Mpc$^3$~\cite{Murase:2016gly,Kowalski:2014zda}, where $E_\nu {\rm d}L_\nu/{\rm d}E_\nu$ is the luminosity per logarithmic neutrino energy bin. Limiting the distance to $4.4$~Mpc, corresponding to $z=10^{-3}$, and requiring to have at least $\sim100$ sources to be compatible with the non-observation of point sources in IceCube data dictates $\rho_0 \gtrsim0.3~{\rm Mpc}^{-3}$ which leads to $E_\nu {\rm d}L_\nu/{\rm d}E_\nu\lesssim2\times10^{36}$~erg/s. While compatible with the constraints on $\rho_0$ from the lack of observed multiplets in IceCube data (which requires $\rho_0 \gtrsim 10^{-5}~{\rm Mpc}^{-3}$~\cite{Lipari:2008zf,Kowalski:2014zda,Murase:2016gly,Capel:2020agp}). and HAWC observation of only a handful of point sources (requiring either $\rho_0 \lesssim 10^{-8}~{\rm Mpc}^{-3}$ or $\rho_0 \gtrsim 10^{-4}~{\rm Mpc}^{-3}$)~\cite{Taboada:2018gmk}, the main challenge to this academic solution is simply a principle of reality: the required densities basically saturate the overall density of galaxies in the universe~\cite{Conselice_2016}, in turn dominated by dwarf galaxies, much smaller than the Milky Way. This would probably require exotic source explanations and raise more problems than it solves, adding to the aforementioned problem (i). 

In summary, for low-$z$ sources the tension with the EGB data (up to 820 GeV) and for $E_{\rm br}=10$~TeV can be eased if the sources are within  $z\sim10^{-3}$, while the tension for $E_{\rm br}=1$~TeV persists as for all the other distributions discussed in previous sections. However, independent arguments point to the unfeasibility of a physically viable extragalactic population which only exists within $\sim4$~Mpc distance. Obviously, shrinking the distance even further is equivalent to requiring a Galactic disk or Galactic halo population as the source of IceCube neutrinos, a possibility on which we comment below, in our final section.

\section{\label{concl}Discussion and conclusions}

The past decade has witnessed the discovery of a high-energy astrophysical neutrino flux by IceCube. Identifying its origin, however, has proved rather challenging, to the point that seven years later there is still no consensus on the sources responsible for it. 

A multimessenger approach can provide important clues in this quest. In particular, gamma-ray data and specifically the EGB set some non-trivial constraints on any putative source candidate. We have revisited this relation, having in mind a specific question: {\it Based on the state-of-the-art knowledge of the classes of objects contributing to the EGB and their lack of correlations with IceCube events, what is the room left by the EGB spectrum for a generic extragalactic class of sources for neutrinos, so that the contribution to the EGB via cascading gamma-rays does not overshoot the data?}

Our results show that the well-known tension for a ``standard'' redshift evolution following SFR history is actually a rather generic result: it holds also accounting for known systematics in the EGB extraction, a more conservative way to account for the error budget of known astrophysical contributions, as well as for different redshift evolutions: not only a BL Lac evolution, but any class of extragalactic sources extending over a large $z$-interval seems in tension with the data, as summarized in Table~\ref{tab:chi2}. We have identified and classified a number of ways to reduce the tension: i) First of all, one must assume that the neutrino flux sharply falls below $\sim 10\,$TeV, i.e. just below the reported ``optimal'' analysis cut. ii) If limiting the EGB analysis only to $E_\gamma\gtrsim 10$~GeV, the tension can be reduced. iii) The tension may be further reduced if sources are pushed to large redshifts or to within the Local Group of galaxies.  

\begin{table}[t!]
\centering
\scriptsize
\begin{tabular}{|c|c|c|c|c|c|c|}
\hline
& & \textbf{SFR} & \textbf{BL LAC} & $z\gtrsim3$ & $0\leq z \leq10^{-2}$ & $0\leq z \leq10^{-3}$ \\
\hline
\multirow{2}{*}{$E_{\rm br}=1$~TeV} & EGB $>100$~MeV & $47$ & $137$ & $35$ & $232$ & $110$ \\ \cline{2-7}
    & EGB $>10$~GeV & $37$ & $113$ & $29$ & $211$ & $103$ \\
    \hline
\multirow{2}{*}{$E_{\rm br}=10$~TeV} & EGB $>100$~MeV & $19$ & $39$ & $4.5$ & $49$ & $5.1$ \\ \cline{2-7}
    & EGB $>10$~GeV & $6$ & $26$ & $3.5$ & $40$ & $4.1$ \\
    \hline 
\end{tabular}
\caption{\label{tab:chi2}The resulted $\Delta\chi^2$ from the fit of EGB data after including the cascaded flux for various cosmic evolution functions $\mathcal{F}(z)$. For the SFR and BL Lac evolution functions see Figure~\ref{fig:fz}. The $z\gtrsim3$ column is valid for any evolution function. For the last two columns we assumed flat distribution within the indicated redshift range. For all the cases the neutrino flux is fixed to the best-fit point of IceCube's Cascade data set $\Phi_{\rm astro}=1.66$ and $s_h=2.53$. For each $E_{\rm br}$, two ranges of EGB data have been considered in the fit, see section~\ref{sec:rob}. To set the scale, $\Delta\chi^2=4.61,6.18,11.83$ means exclusion by $1\sigma,2\sigma,3\sigma$ C.L., respectively.}    
\end{table}

These considerations highlight the importance of a few diagnostics channels to be closely scrutinized in the coming years: a) Refine the IceCube analysis, with a particular attention to the low-energy extent of the spectrum. Even a factor $\sim 2-3$ extension of the flux to lower energies would jeopardize the viability of any conventional way out considered in this article. More accurate measurement of $s_h$, which can be achieved by higher statistics at IceCube-Gen2~\cite{Aartsen:2014njl} in the near future, could also clarify the situation. b) Assess the extent to which alternative components (other than the current radio-galaxy and starburst models) for the EGB below few GeV are viable. Contrarily to the Blazar contribution which is rather well understood and modeled, the current uncertainties are such that surprises cannot be excluded. c) Surveys of the TeV-PeV sky (HAWC and the forthcoming LHAASO~\cite{Cao:2010zz,Bai:2019khm}) are also useful to further constrain a relatively nearby origin. d) From the model-building point of view, it is for instance of interest to explore if energetically viable classes of high-$z$ emitters can be found.

It is possible that progress in one or several of the above lines will lower the tension. Still, even a relatively conventional solution may constitute a rather non-trivial spin-off of the birth of neutrino astronomy. For instance, it would be quite remarkable if IceCube data were to suggest alternative models of the GeV energy range of the EGB!

As we have argued in the main text, however, further contributions to the EGB (gamma's of leptonic origin, byproducts of ultra-high energy cosmic ray propagation and energy losses) rather suggest that the tension may be more severe than what conservatively has estimated here. In that case, the solution may be found in one or several of the following: 

I. A sizable Galactic disk or Galactic halo contribution to the IceCube data at low energy, either astrophysical or exotic. However, conventional Galactic sources would produce a signal peaking in the Galactic plane, which is inconsistent with data~\cite{Aartsen:2017ujz}.  Even an unconventional astrophysical halo source term~\cite{Neronov:2018ibl} seems to be in tension with the limits from KASCADE~\cite{Schatz:2003aw} and especially HAWC~\cite{Harding:2019tez}. Additional arguments suggest that rather hard spectra are required to avoid constraints~\cite{Murase:2013rfa}, harder than those suggested by Galactic cosmic ray models and nowadays, neutrino observations themselves. Finally, even some dark matter decay models~\cite{Esmaili:2013gha,Feldstein:2013kka} which would naturally produce a halo signal are in tension with some of the IceCube data sets~\cite{Cohen:2016uyg}, although viable fits still exist~\cite{Bhattacharya:2019ucd,Harding:2019tez,Chianese:2016kpu,Esmaili:2015xpa,Sui:2018iwo}. All in all, this scenario requires rather peculiar spectral features of this flux, which is qualitatively easier to envisage within exotic models than in astrophysical ones. Further studies of the EGB in Fermi-LAT data at high energy and forthcoming studies of the diffuse flux in HAWC and LHAASO would provide a definite test of this possibility. Note how in this case neutrino astronomy would have anticipated a discovery in a yet unexplored window, the hundreds of TeV-PeV Galactic gamma-ray astronomy.

II. Attribute the bulk of the events to opaque sources, i.e. a new class of objects apparently shining at high-energy only in neutrinos. Some possibilities put forward are low-luminosity {\it choked} gamma-ray bursts~\cite{Murase:2013ffa,Senno:2015tsn}, or hidden AGN cores~\cite{Kimura:2014jba,Murase:2019vdl}. Actually, by {\it opaque sources} one only means that highly energetic gamma rays are damped, and that electromagnetic counterparts of these objects are to be found at lower energies. This case, while consistent with observational constraints~\cite{Murase:2013rfa}, lacks at the moment an independent smoking-gun confirmation, despite a number of searches (see for instance~\cite{Senno:2017vtd,Esmaili:2018wnv}) but can lead to interesting signatures in the X-ray and soft gamma ray bands. Some strategies for such searches are being devised, see e.g.~\cite{Bradascio:2019xdd,Santander:2019yeo}. In this case, neutrino astronomy would be the only direct probe of deeply hidden environments, as the vicinity of supermassive black holes. 
 
III. Not excluding the previous two, it is also possible that the IceCube flux has a complicated multi-component nature, so that linking e.g. the flux at 10 TeV to the flux at PeV scales is {\it not} justified, and multiple components must then be found. The identification of a Galactic contribution only present at low-energy, for instance, would point in this direction. Future water-based neutrino telescopes in the Northern Hemisphere, KM3NeT~\cite{Adrian-Martinez:2016fdl}, would provide important clues in this respect.

As is often the case, a new discovery has brought more questions than answers. Although we cannot predict which solution has been chosen by Nature, the brief overview given above strongly suggests that whatever the solution to this tension is, it is likely going to be very exciting for high-energy astroparticle physics!

\begin{acknowledgments}
We thank K. Murase for comments on an earlier version of this manuscript. 
A. C. thanks the support received by the FAPERJ scholarship No. E-26/201.794/2019. This  work  has been partially supported  by  Univ.  Savoie Mont Blanc,  {\it appel  \`a  projets} ``DIGHESE''. 
\end{acknowledgments}

\appendix

\section{\label{sec:app}EGB data}

For convenience, the EGB data for different foreground modelings and the upper and lower uncertainties (statistical, systematic and Galactic foreground modeling uncertainties added in quadrature) are displayed in Table~\ref{tab:egbdata}.

\begin{landscape}
\begin{table}[h!]
\centering
\scriptsize
\begin{tabular}{|c|c|c|c|c|c|c|c|c|c|c|}
\hline
\multicolumn{2}{|c|}{\textbf{Energy bins {[}GeV{]}}} & \multicolumn{3}{c|}{\textbf{EGB Intensities [cm$^{-2}$ s$^{-1}$ sr$^{-1}$]}}            & \multicolumn{3}{c|}{\textbf{Upper uncertainties (stat. + syst. + FG.)}} & \multicolumn{3}{c|}{\textbf{Lower uncertainties (stat. + syst. + FG.)}} \\ \hline
\multicolumn{1}{|c|}{Lower bound}      & Upper bound     & \multicolumn{1}{c|}{FG Model A} & \multicolumn{1}{c|}{FG Model B} & FG Model C              & \multicolumn{1}{c|}{FG Model A}      & \multicolumn{1}{c|}{FG Model B}      & FG Model C & \multicolumn{1}{c|}{FG Model A}      & \multicolumn{1}{c|}{FG Model B}      & FG Model C                   \\ \hline
0.1                                    & 0.1414          & $3.674 \times 10^{-6}$          & $3.941 \times 10^{-6}$          & $3.339 \times 10^{-6}$  & $6.331 \times 10^{-7}$               & $5.792 \times 10^{-7}$               & $8.190 \times 10^{-7}$       & $1.300 \times 10^{-6}$ & $1.548 \times 10^{-6}$ & $1.004 \times 10^{-6}$\\ \hline
0.1414                                 & 0.2             & $2.321 \times 10^{-6}$          & $2.414 \times 10^{-6}$          & $2.100 \times 10^{-6}$  & $4.530 \times 10^{-7}$               & $4.444 \times 10^{-7}$               & $5.386 \times 10^{-7}$       & $8.768 \times 10^{-7}$ & $9.604 \times 10^{-7}$ & $6.920 \times 10^{-7}$\\ \hline
0.2                                    & 0.2828          & $1.469 \times 10^{-6}$          & $1.590 \times 10^{-6}$          & $1.311 \times 10^{-6}$  & $3.373 \times 10^{-7}$               & $3.160 \times 10^{-7}$               & $4.175 \times 10^{-7}$       & $5.794 \times 10^{-7}$ & $6.857 \times 10^{-7}$ & $4.532 \times 10^{-7}$\\ \hline
0.2828                                 & 0.4             & $9.697\times 10^{-7}$           & $1.091 \times 10^{-6}$          & $8.825 \times 10^{-7}$  & $2.364 \times 10^{-7}$               & $2.034 \times 10^{-7}$               & $2.893 \times 10^{-7}$       & $3.786 \times 10^{-7}$ & $4.870 \times 10^{-7}$ & $3.079 \times 10^{-7}$\\ \hline
0.4                                    & 0.5657          & $6.735 \times 10^{-7}$          & $7.664 \times 10^{-7}$          & $6.184 \times 10^{-7}$  & $1.390 \times 10^{-7}$               & $1.041 \times 10^{-7}$               & $1.796 \times 10^{-7}$       & $2.414 \times 10^{-7}$ & $3.285 \times 10^{-7}$ & $1.928 \times 10^{-7}$\\ \hline
0.5657                                 & 0.8             & $4.871 \times 10^{-7}$          & $5.571 \times 10^{-7}$          & $4.523 \times 10^{-7}$  & $8.218 \times 10^{-8}$               & $4.397 \times 10^{-8}$               & $1.126 \times 10^{-7}$       & $1.621 \times 10^{-7}$ & $2.310 \times 10^{-7}$ & $1.289 \times 10^{-7}$\\ \hline
0.8                                    & 1.1314          & $2.990 \times 10^{-7}$          & $3.457 \times 10^{-7}$          & $2.765 \times 10^{-7}$  & $5.267 \times 10^{-8}$               & $2.526 \times 10^{-8}$               & $7.289 \times 10^{-8}$       & $9.934 \times 10^{-8}$ & $1.456 \times 10^{-7}$ & $7.768 \times 10^{-8}$\\ \hline
1.1314                                 & 1.6             & $1.786  \times 10^{-7}$         & $2.152 \times 10^{-7}$          & $1.690 \times 10^{-7}$  & $3.946 \times 10^{-8}$               & $1.582 \times 10^{-8}$               & $4.829 \times 10^{-8}$       & $5.752 \times 10^{-8}$ & $9.365 \times 10^{-8}$ & $4.822 \times 10^{-8}$\\ \hline
1.6                                    & 2.2627          & $1.089 \times 10^{-7}$          & $1.364 \times 10^{-7}$          & $1.055\times 10^{-7}$   & $2.918 \times 10^{-8}$               & $1.059 \times 10^{-8}$               & $3.226 \times 10^{-8}$       & $3.445 \times 10^{-8}$ & $6.157 \times 10^{-8}$ & $3.123 \times 10^{-8}$\\ \hline
2.2627                                 & 3.2             & $6.932 \times 10^{-8}$          & $8.866 \times 10^{-8}$          & $6.950 \times 10^{-8}$  & $2.009 \times 10^{-8}$               & $6.258 \times 10^{-9}$               & $1.984 \times 10^{-8}$       & $2.142 \times 10^{-8}$ & $4.062 \times 10^{-8}$ & $2.164 \times 10^{-8}$\\ \hline
3.2                                    & 4.5255          & $4.207 \times 10^{-8}$          & $5.494 \times 10^{-8}$          & $4.318 \times 10^{-8}$  & $1.342 \times 10^{-8}$               & $4.371 \times 10^{-9}$               & $1.232 \times 10^{-8}$       & $1.264 \times 10^{-8}$ & $2.536 \times 10^{-8}$ & $1.374 \times 10^{-8}$\\ \hline
4.5255                                 & 6.4             & $2.618 \times 10^{-8}$          & $3.445 \times 10^{-8}$          & $2.730 \times 10^{-8}$  & $8.737 \times 10^{-9}$               & $3.188 \times 10^{-9}$               & $7.667 \times 10^{-9}$       & $7.888 \times 10^{-9}$ & $1.600 \times 10^{-8}$ & $8.983 \times 10^{-9}$\\ \hline
6.4                                    & 9.051           & $1.692 \times 10^{-8}$          & $2.228 \times 10^{-8}$          & $1.781 \times 10^{-8}$  & $5.649 \times 10^{-9}$               & $2.061 \times 10^{-9}$               & $4.809 \times 10^{-9}$       & $5.077 \times 10^{-9}$ & $1.035 \times 10^{-8}$ & $5.944 \times 10^{-9}$\\ \hline
9.051                                  & 12.8            & $1.203 \times 10^{-8}$          & $1.515 \times 10^{-8}$          & $1.260 \times 10^{-8}$  & $3.317 \times 10^{-9}$               & $1.347 \times 10^{-9}$               & $2.790 \times 10^{-9}$       & $3.533 \times 10^{-9}$ & $6.605 \times 10^{-9}$ & $4.095 \times 10^{-9}$\\ \hline
12.8                                   & 18.1019         & $6.754 \times 10^{-9}$          & $8.726 \times 10^{-9}$          & $7.342 \times 10^{-9}$  & $2.074 \times 10^{-9}$               & $7.714 \times 10^{-10}$              & $1.534 \times 10^{-9}$       & $1.958 \times 10^{-9}$ & $3.906 \times 10^{-9}$ & $2.537 \times 10^{-9}$\\ \hline
18.1019                                & 25.6            & $4.376 \times 10^{-9}$          & $5.532 \times 10^{-9}$          & $4.693 \times 10^{-9}$  & $1.221 \times 10^{-9}$               & $4.769 \times 10^{-10}$              & $9.325 \times 10^{-10}$      & $1.271 \times 10^{-9}$ & $2.414 \times 10^{-9}$ & $1.584 \times 10^{-9}$\\ \hline
25.6                                   & 36.2039         & $2.668 \times 10^{-9}$          & $3.329 \times 10^{-9}$          & $2.830 \times 10^{-9}$  & $7.013 \times 10^{-10}$              & $2.823 \times 10^{-10}$              & $5.537 \times 10^{-10}$      & $7.767 \times 10^{-10}$ & $1.430 \times 10^{-9}$ & $9.370 \times 10^{-10}$\\ \hline
36.2039                                & 51.2            & $1.789 \times 10^{-9}$          & $2.164 \times 10^{-9}$          & $1.873 \times 10^{-9}$  & $4.052 \times 10^{-10}$              & $1.822 \times 10^{-10}$              & $3.312 \times 10^{-10}$      & $5.237 \times 10^{-10}$ & $8.941 \times 10^{-10}$ & $6.064 \times 10^{-10}$\\ \hline
51.2                                   & 72.4077         & $1.093 \times 10^{-9}$          & $1.296 \times 10^{-9}$          & $1.135 \times 10^{-9}$  & $2.245 \times 10^{-10}$              & $1.123 \times 10^{-10}$              & $1.890 \times 10^{-10}$      & $3.222 \times 10^{-10}$ & $5.218 \times 10^{-10}$ & $3.633 \times 10^{-10}$\\ \hline
72.4077                                & 102.4           & $6.183 \times 10^{-10}$         & $7.267 \times 10^{-10}$         & $6.400 \times 10^{-10}$ & $1.236 \times 10^{-10}$              & $6.842 \times 10^{-11}$              & $1.062 \times 10^{-10}$      & $1.832 \times 10^{-10}$ & $2.888 \times 10^{-10}$ & $2.041 \times 10^{-10}$\\ \hline
102.4                                  & 144.8155        & $3.084 \times 10^{-10}$         & $3.637 \times 10^{-10}$         & $3.185 \times 10^{-10}$ & $6.608 \times 10^{-11}$              & $4.045 \times 10^{-11}$              & $5.852 \times 10^{-11}$      & $9.413 \times 10^{-11}$ & $1.470 \times 10^{-10}$ & $1.036 \times 10^{-10}$\\ \hline
144.8155                               & 204.8           & $1.925 \times 10^{-10}$         & $2.200 \times 10^{-10}$         & $1.984 \times 10^{-10}$ & $3.748 \times 10^{-11}$              & $2.783 \times 10^{-11}$              & $3.396 \times 10^{-11}$      & $5.901 \times 10^{-11}$ & $8.459 \times 10^{-11}$ & $6.438 \times 10^{-11}$\\ \hline
204.8                                  & 289.6309        & $8.880 \times 10^{-11}$         & $1.015 \times 10^{-10}$         & $9.198 \times 10^{-11}$ & $2.145 \times 10^{-11}$              & $1.791 \times 10^{-11}$              & $2.002 \times 10^{-11}$      & $2.925 \times 10^{-11}$ & $4.025 \times 10^{-11}$ & $3.188 \times 10^{-11}$\\ \hline
289.6309                               & 409.6           & $6.280 \times 10^{-11}$         & $6.862 \times 10^{-11}$         & $6.496 \times 10^{-11}$ & $1.605 \times 10^{-11}$              & $1.418 \times 10^{-11}$              & $1.530 \times 10^{-11}$      & $2.142 \times 10^{-11}$ & $2.601 \times 10^{-11}$ & $2.310 \times 10^{-11}$\\ \hline
409.6                                  & 579.2619        & $2.053 \times 10^{-11}$         & $2.304 \times 10^{-11}$         & $2.152 \times 10^{-11}$ & $9.515 \times 10^{-12}$              & $9.336 \times 10^{-12}$              & $9.425 \times 10^{-12}$      & $9.986 \times 10^{-12}$ & $1.166 \times 10^{-11}$ & $1.064 \times 10^{-11}$\\ \hline
579.2619                               & 819.2           & $9.663 \times 10^{-12}$         & $9.931 \times 10^{-12}$         & $9.800 \times 10^{-12}$ & $6.448 \times 10^{-12}$              & $6.718 \times 10^{-12}$              & $6.570 \times 10^{-12}$      & $6.654 \times 10^{-12}$ & $6.885 \times 10^{-12}$ & $6.766 \times 10^{-12}$\\ \hline
\end{tabular}
\caption{\label{tab:egbdata}The EGB intensities and uncertainties, in $\left[ {\rm cm}^{-2} {\rm s}^{-1} {\rm sr}^{-1}\right]$, for the three modelings of Galactic foreground. The uncertainties include the statistical, systematic and foreground modeling errors added in quadrature. Data taken from~\cite{Ackermann:2014usa}.}    
\end{table}
\end{landscape}


\bibliographystyle{JHEP}
\bibliography{refs}

\end{document}